\documentclass[10pt]{emulateapj}

\shorttitle{From binaries to multiples I: Data.}
\shortauthors{Tokovinin}

\begin{document}

\title{From binaries to multiples I: Data on F and G dwarfs within 67 pc of the
  Sun.}

\author{Andrei Tokovinin}
\affil{Cerro Tololo Inter-American Observatory, Casilla 603, La Serena, Chile}
\email{atokovinin@ctio.noao.edu}

\begin{abstract}
Data on the multiplicity of F- and G-type dwarf stars within 67\,pc of
the Sun are presented.  This distance-limited sample based on the {\em
  Hipparcos} catalog contains 4847 primary stars (targets) with $0.5 <
V-I_C  < 0.8$ and  is $>90\%$  complete. There  are 2196  known stellar
pairs, some of them belong to 361 hierarchical systems from triples to
quintuples.   Models  of   companion  detection  by  radial  velocity,
astrometric acceleration, direct  resolution, and common proper motion
are developed.   They serve to  compute completeness for  each target,
using the information  on its coverage collected here.   About 80\% of
companions  to the  primary  stars  are detected,  but  the census  of
sub-systems in the secondary components is only about 30\%.  Masses of
binary components  are estimated from their absolute  magnitudes or by
other  methods, the  periods of  wide pairs  are evaluated  from their
projected separations.  A third  of binaries with periods shorter than
$\sim$100\,yr  are  spectroscopic and/or  astrometric  pairs with  yet
unknown  periods  and  mass  ratios.   These  data  are  used  in  the
accompanying Paper  II to  derive unbiased statistics  of hierarchical
multiple systems.
%
%\\ Last update: {\it Nov 28, 2013}
\end{abstract}


%-------------------------------------------------------------
\section{Goals and strategy}
\label{sec:goals}

Statistics  of  stellar  multiple  systems is  important  for  several
reasons, the major one being  probably star formation.  Why some stars
are  born with  stellar  companions and  some  are not?   What is  the
relation  between multiplicity, debris  disks, and  planetary systems?
Are  stellar mergers  in multiple  systems  frequent and  how   they
affect the initial mass  function?  Compared to binaries, hierarchical
multiples with three or more components contain additional information
such as  period ratios, mass  ratios, and relative  orbit orientation.
Extracting and  deciphering this  information will help  to understand
star  formation  and, eventually,  to  predict  statistics of  stellar
systems.  Stellar  hierarchies matter because  they evolve differently
from    simple   binaries,    helping   to    form    close   binaries
\citep{Fabrycky2007}  and  more exotic  objects  like blue  stragglers
\citep{BlueStrugglers}.  Hieracrhical systems cause false positives in
the search of exo-planets \citep{Santerne2013}.

The  goal  of  this  work  is  to  establish  unbiased  statistics  of
hierarchical  stellar systems  (triples,  quadruples, etc).   Previous
studies focused  mostly on binaries  and considered multiples  only in
passing.  While  reaching completeness  for binaries is  difficult, it
becomes  even  more  problematic  for  hierarchies.   As  hierarchical
systems are  less frequent than binaries, their  study requires larger
samples.   For  example,  the  25-pc volume  surveyed  by  \citet{R10}
contains  only  454  targets,  56   of  which  (12\%)  are  triple  or
higher-order  multiples  --  too   few  to  grasp  the  statistics  of
hierarchies.

We extend the horizon of previous multiplicity studies to the distance
of 67\,pc, with 10$\times$  more objects. Solar-type dwarfs are chosen
as  primary  targets, and  a  well  defined  volume-limited sample  is
constructed from the {\it Hipparcos} catalog \citep{HIP1}.  Solar-type
dwarfs are traditionally selected for multiplicity study because stars
of  lower  mass are  faint,  while more  massive  stars  are rare  and
distant.   Although the  knowledge of  multiplicity in  different mass
regimes and  environments is needed,  nearby dwarfs are the  first and
easiest step towards this goal.

Our task  is simplified by the  existence of extensive  data on nearby
stars.   Many targets are  being monitored  for exo-planets  in radial
velocity  (RV),  providing at  the  same  time  strong constraints  on
stellar companions.  Collection of published data (data-mining) is the
cornerstone  of  this study.   It  is  complemented  by small  surveys
designed to  fill the lacking  information.  Instead of  attempting to
observe all $\sim$5000 stars with complementary techniques required to
detect companions over  the full range of periods  and mass ratios, we
explore  specific areas of the parameter  space. Particular attention
is directed  to binaries, trying to  convert them into  triples and to
constrain the  frequency of sub-systems.  Detection  limits of various
techniques  are   quantified  and   used  to  correct   the  remaining
incompleteness.

Many stars in this sample host known exo-planets, more planets will be
discovered in  the future. Here  we focus on {\em  stellar} companions
and mention exo-hosts  only in the notes. Study  of planets in stellar
multiple systems is an interesting research topic \citep{Roell2012},
it will be advanced by this data collection.

In the  accompanying Paper II  we present the statistical  analysis of
stellar hierarchies and  place it in the context  of prior work, while
this  first  part (Paper  I)  contains the  data.   It  begins by  the
definition of the sample in \S\ref{sec:sample}, followed by the review
of  data  sources  and  methods in  \S\ref{sec:data}.   Evaluation  of
detection   completeness  is   covered  in   \S\ref{sec:det}.   Tables
containing  information  on  individual  components  and  systems  are
presented in \S\ref{sec:tables}.  The  paper concludes by the overview
of this data collection in \S\ref{sec:overview}.

%-------------------------------------------------------------
\section{The FG-67 sample}
\label{sec:sample}

%\subsection{Definition}

Targets  for  this survey  are  selected  from  the {\em  Hipparcos-2} catalog
\citep[][hereafter HIP2]{HIP2} by the following criteria.
\begin{enumerate}
\item
Trigonometric parallax  $p_{\rm HIP} > 15$\,mas (within  67\,pc of the
Sun,  distance  modulus   $<4.12^m$).   Targets  with  parallax  error
$>7.5$\,mas are excluded.

\item
Color $0.5  < V-I_C < 0.8$  (this corresponds approximately  to spectral
types from F5V to G6V, masses from 0.85 to 1.5 $M_{\odot}$).

\item
Unevolved,  satisfying the condition  $M_{\rm Hp}  > 9(V-I_C)  - 3.5$,
where $M_{\rm  Hp}$ is the  absolute magnitude in the  {\it Hipparcos}
band  calculated with  $p_{\rm HIP}$.  Subgiants are  included  in the
sample.
\end{enumerate}

\begin{figure}
\plotone{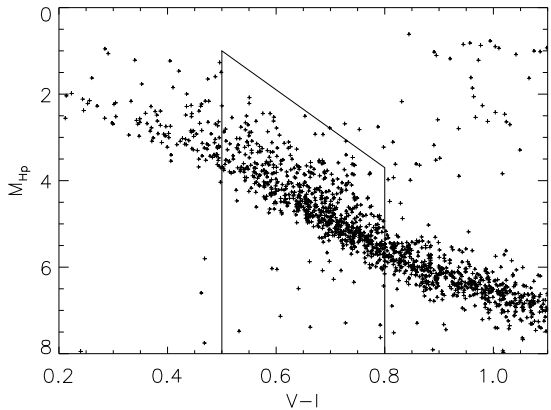}
\caption{Color-magnitude  diagram.   The  lines denote  the  selection
  criteria  of the  FG-67  sample.  Only  {\it  Hipparcos} stars  with
  $p_{\rm HIP} > 30$\,mas are plotted. 
\label{fig:cmd} }
\end{figure}

Figure~\ref{fig:cmd}  shows   the  $(M_{Hp},  V-I_C)$  color-magnitude
diagram (CMD) of  {\it Hipparcos} stars. The upper  cutoff in absolute
magnitude  is set  at about  $2^m$ above  the main  sequence  to avoid
discrimination against multiple stars.  The Hyades cluster in included
in the sample.

\begin{figure}
\plotone{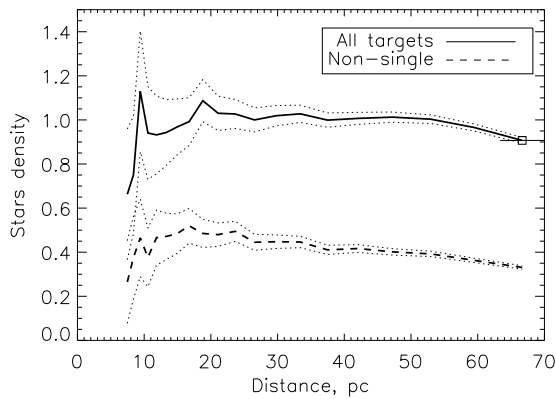}
\caption{Density of FG-67 targets within given distance, normalized by
  0.0043~star~pc$^{-3}$  (thick solid  line).  The  thick  dashed line
  shows the density of  non-single targets.  Dotted lines indicate the
  $\pm  \sigma$ statistical  errors,  the square  with horizontal  bar
  indicates a typical distance error.
\label{fig:checkdist} }
\end{figure}

The criteria  formulated above select  5040 stars from  HIP2. However,
components of  wide binaries with  two individual HIP entries  must be
counted only once (their secondaries are removed from the target list,
even  though  they fulfill  the  sample  criteria).   We also  removed
targets that  have wide  companions more massive  than 1.5\,$M_\odot$.
About 50  other targets (1\%)  are removed for various  other reasons,
e.g.   stars    located   far    below   the   main    sequence   (see
Figure~\ref{fig:cmd}) and stars with  erroneous $V-I_C$ colors in HIP2
(checked  against photometry  in  other bands  and/or spectral  type).
Cleaning reduces  the original selection by 4\%,  leaving 4847 targets
(primary components).

The selection  criteria are blurred by observational  errors in colors
(the  vertical  lines  in  Figure~\ref{fig:cmd}  are  not  sharp)  and
distances.  The actual errors  of {\em Hipparcos} parallaxes sometimes
exceed their formal errors, especially for binaries.  We do not reject
targets  near the 15\,mas  cutoff if  they appear  to be  further away
based on  photometry.  Some solar-mass stars within  67\,pc are missed
in  the  FG-67 sample  because  they were  not  included  in the  {\em
  Hipparcos}  catalog,   for  example  the   nearby  multiple  systems
$\zeta$~Cnc and  $\xi$~UMa.  The masses  of primary components  in the
FG-67 sample are  larger than in the 25-pc sample  of Raghavan et al.;
the lower cutoff  in mass is dictated here by  the completeness of the
{\em Hipparcos} catalog. For this reason, the size of the FG-67 sample
is  less than  the size  of the  25-pc sample  scaled as  cube  of the
distance limit.

Completeness    of    the    FG-67    sample   is    illustrated    in
Figure~\ref{fig:checkdist}. The number  of targets within distance $d$
is proportional to $d^3$, so their spatial density is nearly constant,
dropping only  by 10\%  at 67\,pc. A  drop of  8\% is expected  if the
vertical scale of the Galactic disk is 300\,pc, so the completeness of
the  present sample  is above  90\%.  The  density of  1769 non-single
targets   declines   with   distance   slightly   faster,   indicating
progressively  increasing incompleteness  of the  binary  census.  The
observed multiplicity fraction is  therefore $f_M = 1769/4847 = 0.36$,
to be  compared to the  true $f_M =  0.46$ derived by  \citet{R10} and
confirmed  in  Paper  II.    The  overall  completeness  of  companion
detection is reasonably high, about 80\%.

%-------------------------------------------------------------
\section{Data sources and methods}
\label{sec:data}

\subsection{Data structure}

\begin{figure*}
\epsscale{0.8}
\plotone{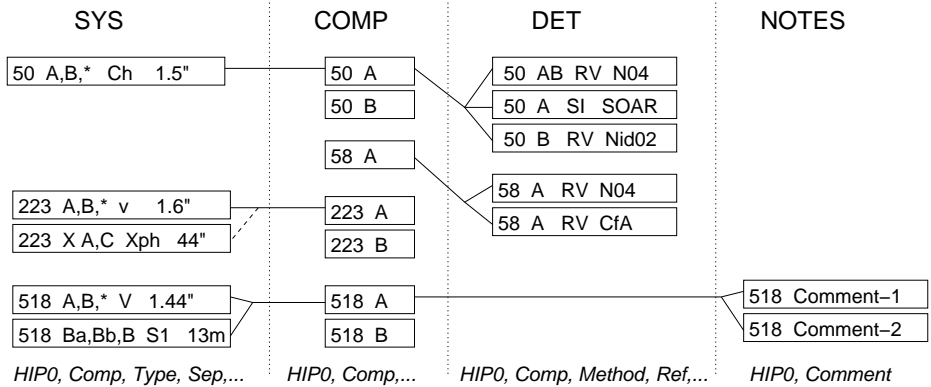}
\caption{Data structure. The four tables SYS, COMP, DET, and NOTES are
  related  by  HIP0,  the   {\em  Hipparcos}  number  of  the  primary
  component.  The  examples show   binary systems  50 and  223 (the
  latter with an optical companion C),  a single star 58, and a triple
  system 518. Each target can  have arbitrary number of records in the
  SYS, DET, and  NOTES tables. 
\label{fig:dat} }
\end{figure*}

All data on a given system are linked by the {\em Hipparcos} number of
its  primary component,  HIP0.  Four  tables, presented  in  detail in
\S\ref{sec:tables}, contain  information on the  individual components
(COMP),  binary  pairings (SYS),  detection  limits  (DET), and  notes
(NOTES),  as  illustrated  in Figure~\ref{fig:dat}.   Custom  software
written in  IDL helps to maintain  this database: browse  and edit the
data, evaluate system  parameters and companion detection probability,
query some catalogs. Components can be placed on the $(V,V-K)$ or $(J,
J-K)$  CMDs to  check the  consistency of  their parallax.   Colors in
various  pass-bands  are  checked   for  consistency  with  colors  of
main-sequence dwarfs.

Each binary system has  two important attributes. The first, component
designation, describes the hierarchy  by a the comma-separated list of
three  components,  (primary, secondary,  parent).   For example,  the
visual binary HIP~518  has components A,B,* (asterisk in  place of the
parent denotes the  {\em root} of the hierarchy,  i.e.  the outer-most
pair).   The spectroscopic  sub-system in  the secondary  component is
Ba,Bb,B.  This  designation is explained  in \citep{Tok06} and  is now
used in the Multiple Star Catalog \citep{MSC}.

The  second  attribute  is  the  {\em type}  of  the  system,  meaning
discovery techniques such as spectroscopic binaries (type 's'), visual
binaries (type  'v'), etc., as  detailed further in this  Section. The
type determines the sense of  system parameters such as separation and
period; they  are either derived  from the orbital solutions  or estimated
(\S~\ref{sec:syspar}). A system can have several types.

\begin{deluxetable*}{l l l}[ht]            
\tabletypesize{\scriptsize} \tablecaption{Bibliographic references and
  their codes
\label{tab:bib}   }      
\tablewidth{0pt} 
\tablehead{ Code   & Reference & Comment, number of targets}
\startdata
2MASS & \cite{2MASS} & $J$, $H$, $K$, companions $5''$ to $30''$ \\
Abt2006 & \cite{Abt2006} & RV (143) \\
ANDICAM & \cite{ANDICAM} & 2MASS companions (66) \\
CfA  & Latham, D. W., 2012, private comm. & RV (1839) \\
Chauvin06 & \cite{Chauvin06} & AO, exo-hosts (17) \\
Chauvin10  &   \cite{Chauvin06}    & AO, young stars (9) \\
Egg2007 & \cite{Egg2007}  &AO, exo-hosts (86) \\
Ginski2012 & \cite{Ginski2012}    & Lucky imaging, exo-hosts (24)\\
Gorynya2013 & Gorynya, N. A., 2014, in preparation & SB orbits (7) \\
Griffin2012 & \cite{Griffin2012}   & SBs in Hyades \\
Halb2012 & \cite{Halb2012}   & RV of CPM pairs (10) \\
Hartkopf2013 & \cite{Hartkopf13} & CPM pairs (18) \\
HIP1 & \cite{HIP1} & Resolved binaries \\
HIP2 & \cite{HIP2} & Position, parallax, PM, $V$, $I_C$ \\ 
Horch2011 & \cite{Horch2011} & Speckle interferometry (9) \\
INT4 & \cite{INT4} & Speckle interferometry and AO \\
Jenkins2010 &  \cite{Jenkins2010}  & AO, exo-hosts (4) \\
Jodar2013   & \cite{Jodar2013} & Low-mass companions (6) \\
Jones2002  &  \cite{Jones2002}  & Precise RV (156)  \\
LAF07  & \cite{LAF07} & AO (20) \\
Lagrange09 & \cite{Lagrange2009} & RV (41) \\
Latham2002 & \cite{Latham2002} & RV (236) \\
LEPINE & \cite{LEP}  & CPM  $(\rho >30'')$ \\
MH09  & \cite{MH09} & AO (122) \\
MK05 & \cite{MK05} & Acceleration binaries \\
MSC & \cite{MSC}  & Multiple systems \\
N04  & \cite{N04} & RV (4080) \\
NICI & \cite{astrom1,astrom2} & AO (107) \\
Nid02 & \cite{Nid02} & Precise RV (438) \\
NOMAD & \cite{NOMAD} & Photometry and PM of secondaries \\
R10 & \cite{R10}  & Stars within 25\,pc  \\
Rameau2013 & \cite{Rameau2013} & AO (12) \\
RoboAO   & \cite{RoboAO} & AO (704) \\
SB9  & \cite{SB9} &  Spectroscopic binaries \\
SEEDS & \cite{SEEDS} & AO (15) \\
SO2011 & \cite{SO2011} & Very wide pairs \\
SOAR & Tokovinin et al. (2010a), other & Speckle interferometry (604) \\
TS02  & \cite{TS02} & RV of visual binaries (104) \\
Tok2006 & \cite{Tok06} & AO (31) \\
Tok2010  & \cite{THH10} & AO (62) \\
Tremko2010 & \cite{Tremko2010} & RV (5) \\
VB6 & \cite{VB6}  & Visual orbits \\
WDS  & \cite{WDS}   & Visual \& CPM companions \\
WSI  & Mason, B. D., 2009, private comm. & Speckle interferometry (1723) 
\enddata                
\end{deluxetable*}

\subsection{Bibliographic references and their codes}

Periods of binary systems span a huge range, from fraction of a day to
Myrs.   To  reach   completeness,  combination  of  various  observing
techniques  and  data sources  is  mandatory.   This  work takes  full
advantage of  extensive data on nearby stars  collected by generations
of  astronomers.  To  a large  extent  it relies  on compilations  and
catalogs.   Instead of giving  proper credit  to the  original authors
(which  would  require  several  thousand  references),  we  cite  the
catalogs whenever  possible.  References on individual  objects can be
obtained from  SIMBAD.  This  data collection is  reasonably complete,
but not free from  omissions. We checked bibliographic references only
for a subset of targets.  Most information was collected by systematic
scanning of major  astronomical journals (up to November  1, 2013) and
complemented  by  some  unpublished  work mentioned  further  in  this
Section.

Table~\ref{tab:bib}  lists  major  sources  used in  this  survey,  in
alphabetic order of the reference  codes adopted here. The last column
contains a short comment on the nature of data and, where appropriate,
gives the number of  targets covered.  In addition, stellar parameters
such as effective temperature and abundance for many FG-67 targets can
be found in the  PASTEL catalog \citep{PASTEL}.  Radial velocities and
kinematics are collected in the XHIP \citep{XHIP}.  These compilations
complement the multiplicity data collected here.

\subsection{Standard relations}
\label{sec:rel}

\begin{figure}
\plotone{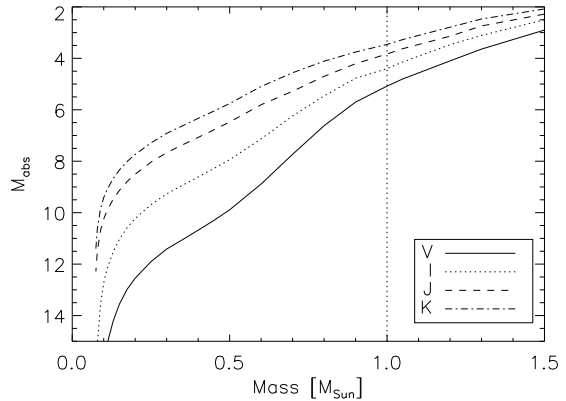}
\caption{Standard   relations  for   main  sequence   stars:  absolute
  magnitude in several photometric bands vs. mass.
\label{fig:absrel} }
\end{figure}

Stars on  the main  sequence show   tight relation  of their  mass with
effective  temperature (hence  spectral type  and color)  and absolute
magnitude.  As  distances to our  targets are known, we  estimate mass
from  absolute magnitude  using standard  relations.  Luminosity  is a
strong function of mass, reducing  the influence of errors in distance
or   photometry  (e.g.   additional   light  from   unresolved  binary
companions) on the estimated masses.

Relations between  mass and absolute  magnitude for the  main sequence
stars  are established  both  empirically \citep{HM93,Delfosse,Lang92}
and through stellar models  \citep{Baraffe,Girardi}.  For dwarfs of $M
< 1 M_\odot$,  the agreement between these sources  is generally good,
on  the order  of  0\fm2  in absolute  magnitude.   We merged  various
standard relations in  a table of absolute magnitudes  vs.  mass for a
grid  of stellar  masses from  0.075\,$M_\odot$ to  2\,$M_\odot$.  For
each mass, the curve $M_{\rm  abs} (k=1/\lambda)$ is almost linear, so
its cubic approximation is good to $\sim 0\fm1$. Here $\lambda$ is the
central  wavelength of  photometric bands,  assumed to  be  [550, 770,
  1250,   2200]\,nm  for   the   $V,I_C,J,K_s$  bands,   respectively.
Polynomial  approximations  allow  us  to  interpolate  standard  relations
$M_{\rm  abs}(\lambda)$ to  other  wavelengths.  They  are plotted  in
Figure~\ref{fig:absrel}.  For a  star of 1\,$M_\odot$, the polynomials
give absolute magnitude  of [5.08,4.40,3.83,3.46] in the $V,I_C,J,K_s$
bands, respectively.

The estimated masses of binary  components are based on their absolute
$V$ magnitudes (code 'v' for the  mass) or, in a few occasions, on the
infrared magnitudes (code 'k').   The values are interpolated linearly
in the  $M_{\rm abs},  M$ table.  We  compared masses of  single stars
with masses  estimated by \citet{Casagrande}  from evolutionary tracks
and found  a good correspondence.  However, the  masses estimated here
are on  average 9\%  larger. Subgiants are  brighter and  more massive
than main-sequence  stars, they  extend the upper  mass limit  of this
sample to  $\sim 1.7\,M_\odot$. There are  63 targets with  $M > 1.5\,
M_\odot$  and 10  targets with  $M  < 0.85\,  M_\odot$, the  remaining
98.5\% have masses within these limits. The median mass of the primary
targets is $1.14\,M_\odot$.

The  following sub-sections  review data  on binaries  by the  type of
their discovery technique, in order of increasing period.

\subsection{Spectroscopic binaries (S1,S2,s)}

Most spectroscopic binaries (SBs) with known orbits are retrieved from
the   on-line  SB9   catalog  \citep{SB9},   complemented   by  recent
publications where necessary.   Single- and double-lined binaries have
types  'S1'  and 'S2',  respectively.   The  masses  of the  secondary
components are derived either from the known mass ratio in the case of
S2  (mass code  'q')  or as  a  minimum  mass  for orbital  inclination  of
$90^\circ$  inferred from  the mass function and the mass of
the primary component (mass code 'm').

A  large  fraction  of  our  sample  (84\%)  was  surveyed  in  RV  by
\citet{N04}.   For 261  stars,  variable RV  was  detected, but  their
orbits are not known. Such cases  are coded by 's' in the system type.
When double lines  were seen in several spectra  (type 's2'), the mass
ratios were  derived, while the orbital period  still remains unknown.
Preliminary  orbital  solutions for  many  binaries  were obtained  at
Center  for  Astrophysics  (CfA)  by  D.  W.   Latham  (2012,  private
communication), extending the  survey by \citet{Latham2002}.  In some
cases,  only the orbital  period is  known to  the author,  leaving the
minimum secondary  mass undetermined.  For  double-lined binaries with
known  period and  unknown RV  amplitudes, the  mass ratio  of  0.8 is
assumed (this is the median value  for all S2), and the secondary mass
code in this case is  'e' (estimated).  Orbits of several short-period
binaries  discovered  by  \citet{N04}  and  not covered  by  CfA  were
recently determined by N.~Gorynya (2014, in preparation).

The  large  volume  of  precise  RV  data  accumulated  in  search  of
exo-planets remains, for the  most part, unpublished and inaccessible,
with a few exceptions \citep{Nid02,Jones2002}.  \citet{TS02} monitored
RVs of visual binaries  to characterize the frequency of spectroscopic
sub-systems, \citet{Halb2012} studied RVs of wide CPM pairs.  The work
of  \citet{Griffin2012}   demonstrates  the  power   of  long-term  RV
monitoring by detecting all SBs in the Hyades.

Some close  binaries are  eclipsing.  They have  type 'E'  (26 total).
Most of them also have  known spectroscopic orbits. Considering the RV
coverage, we did not  search for eclipsing binaries systematically and
do not account for detection  of eclipsing binaries in this study.

\subsection{Astrometric binaries (a,A)}

Binary   stars   are   detectable   from  their   accelerated   motion
\citep{MK05,Frankowski}.  In  some cases, acceleration  was measured by
the {\it Hipparcos} mission over  its 3.2-yr duration; these stars are
known  as  $\dot{\mu}$ binaries  or  G-type  solutions.  In  addition,
accelerated motion in the  so-called $\Delta \mu$ binaries is revealed
by  a significant  difference between  the short-term  {\it Hipparcos}
proper motion (PM) and the long-term PM from the {\it Tycho-2} catalog
exploiting a time base of  almost a century.  Acceleration binaries of
both kinds are coded by type 'a'.  For a fraction of them, astrometric
orbital solutions were derived  in the original {\em Hiparcos} catalog
or later \citep[e.g.][]{GM06}.  In  those cases (type 'A'), the orbital
periods are known.  Combining periods with estimated masses of primary
components and  the distance, we compute the  apparent semi-major axis
$a$ from the  third Kepler law.  The ratio of  the astrometric axis to
$a$ allows us to estimate the mass ratio of A-type binaries (mass code
'q').

So far,  little is known about unresolved  acceleration binaries (some
of those  are also s-type,  i.e. have variable  RV but no  SB orbits).
Yet they cover  an important range of orbital periods from  a few to a
hundred  years where  alternative  detection techniques  are not  very
efficient,  especially for  low-mass  companions.  A  subset of  these
stars  were targeted  by a  dedicated adaptive  optics (AO)  survey at
Gemini-S \citep{astrom1,astrom2}.  About a third were resolved (turned
into type 'v'), allowing estimates of the companion's mass and orbital
period.   The  remaining acceleration  binaries  have companions  too
faint  and/or too  close to  be  resolved.  Some  of those  unresolved
companions could be white dwarfs (WDs).

The Gemini survey revealed new things about acceleration binaries.  It
was  established  that   some  acceleration  solutions  are  spurious,
resulting  from the  ill-conditioned least-squares  problem.   The new
{\em Hipparcos}  reduction, HIP2, eliminated most  of those solutions,
but missed many real acceleration binaries, thus being of little help.
Some stars with $\dot{\mu}$ accelerations actually are relatively wide
binaries resolved with  AO \citep{astrom2} -- too wide  to explain the
acceleration.  This group is a mixture of spurious accelerations where
astrometric   ``noise''  from  faint   companions  was   amplified  in
ill-conditioned  solutions, and triple  stars where  the accelerations
are produced in the inner sub-systems.

Our simulations  demonstrated that most $\dot{\mu}$  binaries are also
detectable by  the $\Delta  \mu$ method.  Therefore,  in this  work we
consider  only $\Delta  \mu$ astrometric  binaries and  do  not accept
$\dot{\mu}$  binaries  as real  unless  they  are  confirmed by  other
methods or have known astrometric orbits. 

\subsection{Close resolved binaries (v,V)}

In  the resolved  binaries, the  companion  is detected  by its  light,
unlike the RV and acceleration methods where only its gravity matters.

The  {\it  Hipparcos} experiment  provided  companion solutions  (i.e.
resolved visual binaries) with a  more or less uniform detection depth
for all  targets, as needed  for this statistical work.   In addition,
close  pairs resolved  by speckle  interferometry and  adaptive optics
(AO) are  collected in  the INT4 catalog  \citep{INT4}, and  all known
visual and occultation binaries are catalogued by the WDS \citep{WDS}.
These inhomogeneous data come from various sources.

Most visual binaries  with separation under $3''$ are  denoted as type
'v'  and are  assumed to  be physical  systems, considering  the small
probability of  finding a random  (and usually bright) star at  such small
separations.  Whenever visual orbits  are available in the VB6 catalog
\citep{VB6}, the type becomes 'V'  and we list the true orbital period
and  semi-major axis  instead of  separation.  Otherwise,  the orbital
period $P^*$ is estimated from the separation (see \S~\ref{sec:syspar}). 

\subsection{Wide companions (C)}

Many  wide  companions  listed  in  the  WDS  are  chance  projections
(optical), denoted  as type  'X'. This can  be revealed by  their fast
relative motion  incompatible with  a Keplerian orbit, or by companion's
magnitude and color  that do not match the values  expected for a main
sequence  dwarf  at  the same  distance  as  the  primary, or  by  the
difference in RV. When the apparent  motion in a wide pair is caused by
the  PM of its  primary component -- {\em  reflex PM} -- its optical
nature is obvious.  On the  other hand, when the wide companion
is  real  (physical),  it  is  denoted  as  type  'C'  with  following
small-letter  qualifiers  h, m,  p,  r  that  show which  criteria  of
physical  relation are fulfilled:  constant relative  position, common
PM, matching  photometric distance, or matching  RV \citep[see details
  in][]{MSC}. Optical companions may also have these qualifies to show
which criterion  was used for  their rejection.  Optical  systems from
the WDS and other doubtful binaries  are included in the SYS table for
completeness  (e.g.   HIP~223  A,C  in  Figure~\ref{fig:dat}),  but  are
ignored in the statistical analysis.

Binary companions with  separations from $5''$ to $30''$  can be found
in  the  2MASS  point  source  catalog,  with  well-defined  detection
limits. This  is a  valuable complement to  the heterogeneous  data of
WDS,   especially  in  the   low-mass  regime   \citep{ANDICAM}.   The
photometry in  2MASS discriminates against  unrelated (optical) stars,
but only out to moderate separations and in not too crowded fields. At
small  separations  $\rho \la  5''$,  the  2MASS  photometry of  faint
components is  usually distorted by bright  primaries, so case-by-case
checks are  necessary (we ignore close  2MASS pairs which  do not have
additional  evidence   of  their   veracity).   On  the   other  hand,
contamination by the field stars becomes important at $\rho \ga 10''$.
This  is why  second-epoch  imaging was  needed  to confirm  candidate
companions found in 2MASS.   The work, started in \citep{ANDICAM}, was
extended to  $\rho <30''$ (unpublished  results of this  extension are
mentioned  in  the notes  as  ANDICAM2), but  it  does  not cover  the
northern  sky. Physical nature  of some  northern candidates  found in
2MASS could be  confirmed by archival optical images  or PMs. Overall,
43 binaries from 2MASS are added here to the 47 new pairs confirmed in
\citep{ANDICAM}.

Components  of  some wide binaries   have  different  parallaxes  in  {\em
  Hipparcos}, for example 20$\pm$5\,mas and 8$\pm$3\,mas for HIP~76888
and 76891, respectively.  Yet this  is a physical binary STF~1966 with
$23''$ separation  observed for 173\,yr. The  photometric parallax of
both components is slightly less than 15\,mas.  There was a problem in
the  {\em  Hipparcos} data  reduction  for  binaries with  separations
around  $20''$  or  binaries  containing sub-systems,  like  HIP~43947
\citep{THH10}.   In those instances  we adopt  same parallax  for both
components.

At separations  $\rho >  30''$, both photometry  and PM are  needed to
distinguish  true  (physical) companions  from  other stars.   Uniform
screening of  stars within 67\,pc  for wide companions down  to $V=19$
became possible  with the  SUPERBLINK survey \citep{LEP},  except 39\%
targets with small PM. In that work, the probability of each companion
being physical was estimated.  Here  we accept all CPM companions with
probability of $>$50\% and  add some lower-probability candidates with
a high  PM or other indications  that they are  real pairs. Subjective
decisions on the status of some wide companions were thus made.

In addition  to the SUPERBLINK  survey, CPM pairs were  retrieved from
the WDS \citep[including the  recent addition by][]{Hartkopf13} and by
matching the {\em Hipparcos} catalog  entries in PM and parallax.  The
latter method  was also used  by \citet{SO2011} to identify  very wide
co-moving  pairs or  groups  of  stars.  These  pairs  are members  of
kinematic groups rather than bound binaries, hence we do not include
them in  the SYS table,  but mention in  the notes. There is  no clear
distinction between  true (bound) wide binaries and  members of moving
groups  \citep{Caballero2010}.  Members  of  the Hyades  cluster  have
common PM and distance, but are not binaries.

The probability of finding false  CPM companions is larger for targets
with small PM and/or in  crowded fields.  If a substantial fraction of
CPM companions were  optical, we expect such pairs  to have smaller PM
and larger crowding  $N^*$.  Comparison of the median  PM and crowding
for  335  binaries  with  $\rho >30''$  (133\,mas\,yr$^{-1}$  and  16,
respectively)   with    the   medians   for    the   complete   sample
(125\,mas\,yr$^{-1}$ and 22)  shows the opposite trend. Statistically,
the sample  does not  contain false wide  binaries and  might actually
miss some real CPM binaries.

\subsection{Dedicated surveys}

Nearby stars  within 25\,pc  were thoroughly surveyed  by \citet{R10}.
We  include information  from  that work  and  use it  as  a check  of
completeness.  Despite substantial observational effort, parameters of
a  few  nearby  binaries  were   still  undetermined  and  had  to  be
``guessed''. There  are 243 targets (out  of 454) in  common with this
work, the rest have masses  smaller than the FG-67 limit. Interestingly,
Raghavan et al. list  three hierarchical multiples with 5-6 components
(HD~68257, 146361, 186858).   The first is missed here  because it has
no HIP number,  the last is only quadruple (the  RV variability of the
component F=HIP~97222  is questioned).  Despite  the 10$\times$ larger
size of  the present sample, it  contains only 5  known quintuples (of
which just one, HD~146361=HIP~79607, is within 25\,pc) and no sextuple
systems.

Only a small  fraction of nearby dwarfs were targeted  by AO in search
of  companions to exo-planet  hosts \citep{Egg2007},  to spectroscopic
binaries \citep{Tok06}, or  to young stars \citep{MH09,Chauvin10}.  In
the  context  of this  project,  we  surveyed  with AO  wide  binaries
\citep{THH10} and astrometric binaries \citep{astrom1,astrom2}.

Speckle interferometry and  lucky imaging in the visible  do not go as
deep as AO,  but cover a larger number of stars  owing to their better
efficiency.    Many   FG-67  targets   were   observed  with   speckle
interferometry   at   the   SOAR  telescope   \citep[][and   following
  papers]{TMH10},   discovering  new   close  sub-systems   in  visual
binaries,  resolving  some acceleration  binaries,  and following  the
orbital motion of ``fast'' close pairs.  Some unpublished speckle data
on G-dwarfs  were provided  by B.~Mason (private  communication), they
are referenced  as WSI (Washington Speckle Interferometry)  in the DET
table.

Almost 600 stars  from the FG-67 were recently  targeted by the RoboAO
system  at  the 1.5-m  Palomar  telescope  \citep{RoboAO}.  This  work
focused  on resolving  faint secondary  components to  constrain their
poorly  known  multiplicity. In  addition,  many  close binaries  were
observed to look for more distant and faint tertiary companions.

\begin{deluxetable}{l c c l}[ht]            
\tabletypesize{\scriptsize} \tablecaption{System types and parameters
\label{tab:types}   }      
\tablewidth{0pt} 
\tablehead{Type   & Sep. & Per. & $M_2$ code}
\startdata
Spectroscopic (S1,S2) & $a$ & $P$ & m,q \\ 
Spectroscopic (s)     & 0   & 0   & -   \\
Astrometric (A)       & $a$ & $P$ & q \\
Acceleration (a)      & 0   & 0   & -   \\
Visual (V)            & $a$ & $P$ & v,k \\
Visual (v)            & $\rho$ & $P^*$ & v,k \\
Wide  (C)             & $\rho$ & $P^*$ & v,k \\
Optical (X)           & $\rho$ & 0 & - 
\enddata                
\end{deluxetable}       

\subsection{Estimation of binary parameters}
\label{sec:syspar}

As mentioned above, the exact  meaning of separation and period in the
SYS table  depends on the system  type.  These parameters,  as well as
masses,  are estimated  automatically  by a  recursive algorithm  that
takes care of  sub-systems. For each target (HIP0  number), it selects
the  data on  all  its systems  and  finds the  outermost binary,  the
root. Each  component of  this binary is  checked for the  presence of
sub-systems,  using the parent  name in  the system  designation.  The
sub-system,  when present,  is evaluated  first  and the  mass sum  is
assigned to its parent with a code 's' (sum). 

For systems  of types S1,  S2, A, and  V, the true orbital  periods are
known.  In  addition, the V-type  systems have known  semi-major axis,
only the masses of both components need  to be evaluated.  Cases where the mass
sum inferred from the  visual orbit and parallax differs significantly
from the  estimated sum of  component's masses are  commented, usually
indicating  poor quality of the visual orbit.

For resolved  systems with  unknown periods (types  v, C),  a probable
orbital period $P^*$ is found by the third Kepler law by assuming that
the angular separation $\rho$ equals orbital semi-major axis $a$,
\begin{equation}
P^* =  [\rho^3 \,  p_{\rm HIP}^{-3} (M_1  + M_2)]^{1/2}, 
\label{eq:P*}
\end{equation}
where $  p_{\rm HIP}$ is  the parallax of  the main target,  $M_1$ and
$M_2$  are masses of components  in units  of solar  mass, and  $P^*$ is
the estimated period  in years.  Simulations  of binary stars  with random
orbital phases and eccentricities  show that the median ratio $\rho/a$
is  indeed  close  to  one,  depending slightly  on  the  eccentricity
distribution.  For a cosine $e$-distribution with $\langle e \rangle =
0.5$,  the 10\%, 50\%,  and 90\%  quantiles of  the $\rho/a$  ratio are
0.38,  0.90,  1.42.   If  the  eccentricity  distribution  is  linear,
$f(e)=2e$,  these  quantiles  are  0.37,  0.98,  1.59.   In  any  case,
$\rho/a<2$, so  the ratio of  estimated and true periods  $P^*/P$ does
not  exceed $2^{3/2} =  3.17$.  By  assuming $a  = \rho$,  we estimate
orbital periods to within a factor of 3, typically.

For the  unresolved binaries with known  period (types S1,  S2, A), we
use the same formula to estimate the orbital semi-major axis, which is
listed in place of the separation. The separation of resolved pairs is
not replaced by those estimates,  however. The value of separation for
resolved  binaries with  detectable  motion is  ambiguous, usually  it
corresponds to  the latest  measured separation listed  in the  WDS or
INT4.

The program  that computes binary periods or  separations also assigns
masses to the  components. When the visual magnitudes  are listed, the
masses are evaluated from  the absolute $V$-magnitudes and receive the
code 'v'.   When the magnitude difference  of a binary  is measured at
some  wavelength and  the total  combined $V$  magnitude is  known, we
solve  for component  masses  that  match those  two  numbers and  the
distance, using standard  relations.  For systems of types  S1 and S2,
the secondary mass has codes 'm' (minimum) and 'q', respectively.  The
mass code 'q' is also  assigned to the secondary components of systems
with astrometric  orbits (type A).   Finally, the code 'r'  means that
the component's masses are taken from the literature; these masses are
not overridden by any of the above estimates.

A summary of system types, corresponding meaning of the separation and
period in the SYS table, and  codes of secondary masses is provided in
Table~\ref{tab:types}.   Spectroscopic and acceleration  binaries with
unknown  orbits are  the worst  case: we  do not  know  their periods,
separations, and  mass of the  secondary components. All  these fields
have the default zero values in the SYS table.

\subsection{False and dubious binaries}
\label{sec:false}

Some  stars declared  to be  binaries  are in  fact single.   Spurious
discoveries  are  produced by  all  techniques  discussed above.   For
example, one slightly deviant RV  measure can cause  a formal detection
of  RV variability  of  a  single star.   Some  accelerations in  {\em
  Hipparcos} are spurious. Visual observers and speckle interferometry
alike  produced a  number  of  false resolutions  of  single stars  or
sub-systems,  sometimes   with  several  ``confirming''  measurements.
Finally,  the physical  nature of  some wide  CPM pairs  is uncertain,
especially   when  the  PM is  small  and  the field  is
crowded. On the other hand,  apparent motion of a rejected wide binary,
deduced from  the first  and last observations  listed in the  WDS, can
appear too fast simply because the measurements are inaccurate.

In this  work, decisions  on accepting or  rejecting binary  pairs are
taken on the  basis of available data which  are not always conclusive.
Such cases  are marked by the  question mark in the  system type. There
are 346 question marks, about 10\%  of all systems.  Of those, 115 are
type  'a'   (rejected  accelerations),  75  type   's'  (uncertain  RV
variability),  99 type  'C' (uncertain  CPM  pairs), and  35 type  'v'
(spurious  resolutions).  Some  subjectivity  is  therefore
unavoidable in this work.  The actual proportion of wrongly accepted or
rejected  binary  pairs should  be  much less  than  10\%,  but it  is
difficult to evaluate.

\subsection{White dwarf companions}
\label{sec:WD}

Some FG-type dwarfs were originally paired to more massive stars which
evolved and became white dwarfs (WDs).  Those ``Sirius-like'' binaries
where  WD is  paired  to a  main-sequence  star \citep{Holberg13}  are
difficult to detect. Holberg et al. estimate the density of such pairs
in the  20-pc volume  as $3.3\times 10^{-4}$\,pc$^{-3}$ and  their fraction
among main sequence  stars of less than 1.2\%.   His table contains 15
WD binaries with components of spectral types F5V to G6V located within 67\,pc
(3 of  those are not  in this sample  because they are not  present in
{\em Hipparcos}).  Three of those binaries are within 20\,pc, giving a
rather uncertain estimate  of WD fraction in the  FG-67 sample as 2\%.
This fraction rises to 4\% if we take the above density of Sirius-like
binaries,  suppose that  half are  paired to  the F-  or  G-stars, and
compare    with    the     spatial    density    of    our    targets,
$4.3\times 10^{-3}$\,pc$^{-3}$  (Figure~\ref{fig:checkdist}).  Actually, 22
binaries are known or suspected to contain WD companions (HIP numbers:
11028, 18824, 20284, 27878,  29788, 32329, 37853, 54530, 60081, 64150,
77358,  80182,  80337,  81478,  81726, 83431,  95293,  99956,  103735,
104101,  113231, 118010), 14  of them  with separations  above $30''$.
The  majority  of Sirius-like  binaries  in  the  FG-67 sample  remain
undetected, but some may hide among the acceleration binaries.

\begin{deluxetable*}{l cc cccc  cccc l} 
\tabletypesize{\scriptsize}      
\tablecaption{Representative detection limits for resolved binaries
\label{tab:lim}   }     
\tablewidth{0pt} 
\tablehead{Method & $N$ & $\lambda $ & $\rho_1$& $\rho_2$& $\rho_3$&$\rho_4$&  
$\Delta m_1$&  $\Delta m_2$ &$\Delta m_3$ &$\Delta m_4$& Ref. \\
(1) & (2) & (3) & (4) &  (5) & (6)& (7) & (8) & (9) & (10) & (11) & (12)  }
\startdata
HIP1 & all & 550 &  0.09 & 0.14 &  0.4 & 10.0 &  0.0  & 2.2  &   4.0  &  4.3  & \citet{HIP1} \\
WDS & all  &  550 &  0.15 & 1.0  & 10.0  & 30.0 & 2.5  & 5.0  & 8.0    & 9.4  & $5 + 3 \log_{10} \rho ''$ \\    
Speckle &  604 &   540 &  0.03 & 0.15 &  1.00 & 1.50 &  0.50 & 4.33 &   5.63 &  5.63 & Tokovinin et al. (2010a) \\
AO &  122 &  2200 &  0.09 & 1.00 &  2.00 & 5.00 &  1.00 & 6.32 &   8.60 &  9.90 & \citet{MH09} \\
AO &  107 &  2272 &  0.054& 0.15 &  0.90 & 9.00 &  0.0  & 5.42 &   7.48 &  7.48 & \citet{astrom1,astrom2} \\
2MASS & all & 2200&  3.0  & 9.0  &  20.0 & 30.0 &  0.0  & 7.5  &   10   & 11 & \citet{ANDICAM}     
\enddata                                                                                                                       
\end{deluxetable*}                                                                                              

%-------------------------------------------------------------
\section{Detection limits}
\label{sec:det}

Knowledge of the probability of companion's detection as a function of
its orbital period  $P$ and mass ratio $q =  M_2/M_1$ is necessary for
deriving  unbiased multiplicity  statistics.   Corrections for  missed
companions are  larger for triples  and quadruples than  for binaries.
If  we detect companions  with a  probability of  0.8, the  chance of
discovering a triple  system (two companions) is $0.8^2  = 0.64$.  It
can be even  less because detection of sub-systems  in the {\em secondary}
components  of  known   binaries  meets  with  additional  difficulty.
\citet{R10} note  that improved completeness  of their survey  did not
change the binary fraction  derived in earlier works \citep{DM91}, but
doubled the number of known hierarchies within 25\,pc.

In  this  Section,  we   present  algorithms  used  to  translate  the
observational coverage of  each target or component (as  listed in the
DET  table) to  the  detection probability  in  the $(P,q)$  parameter
space.  The algorithm is rather  straightforward in the case of direct
resolution  (AO imaging  and  speckle interferometry).   We apply  the
default resolution  limits of  {\em Hipparcos} and  2MASS to  all main
targets  and to  some  secondary components.   In  addition, the  {\em
  Hipparcos} acceleration  $\Delta \mu$ detects  binaries with periods
from few  to hundred  years with a  probability that is  determined by
simulations  in  \S~\ref{sec:detacc}.  At  shorter  periods, the  main
discovery technique  is spectroscopy, where  the detection probability
is also found by simulation and related to the number of observations,
their precision, and time coverage (\S~\ref{sec:detrv}).

Resolution of a binary constrains,  to some extent, sub-systems in its
secondary component.   Similarly, RV data on  unresolved visual binary
tell  something  about  potential   sub-systems  in  its  primary  and
secondary  components.   Detection  of  sub-systems in  the  secondary
components  is also  covered in  this Section.  Binarity  of secondary
components was frequently neglected  in previous works on multiplicity
statistics. 

The information on detection limits and their modeling are necessarily
approximate.   Here we  tend  to adopt  optimistic (deeper)  detection
limits, so  that the completeness  correction becomes smaller  and the
estimated  multiplicity becomes  less.  In  other  words, conservative
estimate of multiplicity requires generous allocation of detection
limits.

\subsection{Resolved binaries}
\label{sec:detvis}

Maximum magnitude  difference of binaries resolved  by {\em Hipparcos}
(companion solutions) $\Delta Hp(\rho)$  shows a well-defined limit 
depending on the angular separation $\rho$ ($\Delta Hp <2.2$ at $\rho =
0\farcs14$  and $\Delta  Hp <4$  at $\rho>0\farcs4$).   Similar limits
exist for other imaging  techniques like AO and speckle interferometry
(see Table~\ref{tab:lim}). They are translated to the detection limits
in  the  $(P,q)$ space  in  the following  way.   Each  period $P$  is
converted  to  separation $\rho$ using equation~\ref{eq:P*}.   The  absolute
magnitude of the primary component at wavelength $\lambda$ is computed
from its  mass $M_1$  using the standard  main sequence relation (\S~\ref{sec:rel}).
We  add  the maximum  magnitude  difference  of detectable  companions
$\Delta m(\rho, \lambda)$, convert back into mass $M_{2, {\rm min}}$ with the
same standard relation, and obtain $q_{\rm min} = M_{2, {\rm min}}/M_1$.

\begin{figure}
%\epsscale{0.85}
\plotone{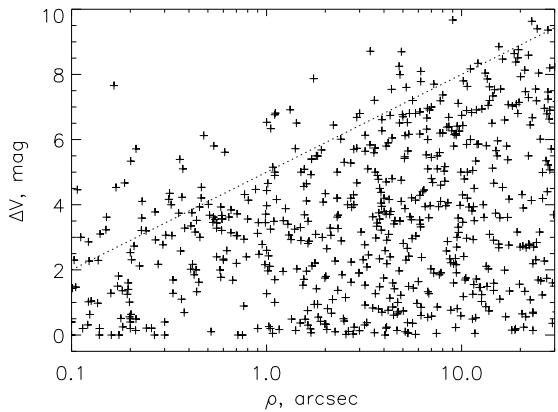}
\caption{Magnitude  difference  vs.  angular separation $\rho$  for  resolved
  binaries  (types v  and  C) in  the  FG-67 sample.  The dotted  line
 is $\Delta V = 5 + 3 \log_{10} (\rho)$.
\label{fig:dV-sep} }
\end{figure}

The   detection limits of imaging  techniques are represented
here by  4 values of separation $\rho_i$,  corresponding $\Delta m_i$,
and  the  imaging wavelength  $\lambda$.   The  $\rho_1$ and  $\rho_4$
define  the  minimum  and   maximum  range  of  surveyed  separations,
respectively, and  $\Delta  m(\rho)$ at intermediate separations is
linearly interpolated between the 4  nodes, describing this curve by 3
linear segments.  Obviously, this is a crude approximation, as well as
the  assumption  that the  probability  of  companion detection  drops
sharply from one to zero  at $q < q_{\rm min}$.  Representative limits
are given in Table~\ref{tab:lim}. It  lists the number $N$ of targets from
the  FG-67 sample covered  by each  work, $\lambda$,  $\rho_i$, median
$\Delta m_i$, and the reference.   We presume that all primary targets
were  examined by  visual observers  and adopt  a  somewhat optimistic
limit 
\begin{equation}
\Delta  V <  5 + 3  \log_{10} (\rho/1'')
\label{eq:WDS}
\end{equation}
which delineates  the upper envelope of the  companion distribution in
the $(\rho, \Delta V)$ plane (dotted line in Figure~\ref{fig:dV-sep}).

\begin{figure}
\epsscale{1.0}
\plotone{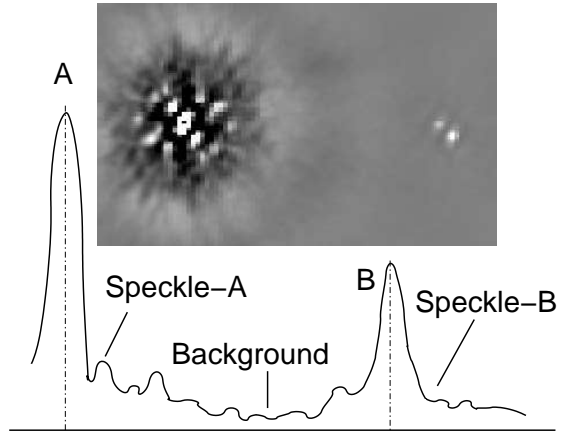}
\caption{Detection  of  sub-systems  in  the secondary  components  of
  resolved  binaries.  The  image  shows detection  of the  sub-system
  Ba,Bb  (0\farcs12,  $\Delta  m  =  0.9$)  in  the  binary  HIP~44874~AB
  (1\farcs79, $\Delta  m = 4.8$) observed  with speckle interferometry
  at SOAR  in 2013.   The curve  illustrates noise in  the image  of a
  binary,  dominated  by  the  residual speckle  structure  near  each
  component and by the background fluctuations away from them.
\label{fig:speckle} }
\end{figure}

Secondary  components targeted  {\em  individually} by  AO or  speckle
interferometry are treated in the same way as the primary targets.  In
addition, observations  of a resolved  binary AB place  constraints on
the  existence   of  resolved  companions   (sub-systems)  around  its
secondary  B.  In  hierarchical multiples,  such sub-systems  can have
separations $\rho  < \rho_{AB}/2$  (otherwise the subsystem  cannot be
attributed  to  B).   The  known  detection  limit  $\Delta  m_A(\rho,
\lambda)$  on companions  around A  is  translated into  the limit  on
companions  around B  by assuming  that both  limits correspond  to $5
\sigma_I$, where the rms intensity fluctuation $\sigma_I(\rho)$ is the
quadratic sum of the residual  speckle noise (which dominates at small
separations) and the  background noise (Figure~\ref{fig:speckle}).  In
the vicinity  of the  secondary, the background  noise and  the scaled
speckle noise  define $\sigma_I$.  This  logic leads to  the following
formula (we omit here its derivation):
\begin{eqnarray}
\lefteqn{
\Delta m_B(\rho)  =   -2.5  \log_{10}  [5  \sigma_I(\rho)/I_B]  } \\
 &&  \approx  -1.25 \log [ 10^{0.8[\Delta m_{AB} - \Delta m_A(\rho_{AB})]}
+ 10^{-0.8 \Delta m_A (\rho)} ], \nonumber 
\label{eq:dm}
\end{eqnarray}
where  $\Delta  m_A(\rho)$  is  the  detection  limit  for  the  primary
component  A, $\rho_{AB}$  is the  separation  of the  AB binary,  and
$\Delta m_{AB}$ is its  magnitude difference. We compute the detection
limits for sub-systems around secondary components of resolved binaries
using this formula.

Many  wide binaries do  not have  any constraints  on the  binarity of
their  secondary  components.   Does  this  mean  that  the  secondary
component  itself  can be,  say,  a  $5''$  pair?  Indeed,  new  close
sub-systems  were  discovered  by  the targeted  survey  of  secondary
components with RoboAO \citep{RoboAO}.  However, an obvious pair would
be  noted in  the optical  or infrared  images even  without dedicated
observations.  We  presume (optimistically) that  the detection limits
of 2MASS apply to the secondary  components with $\rho > 6''$, down to
$K_s  = 16$.   This  assumption constrains  relatively wide  secondary
sub-systems, which are not frequent anyway.

This reasoning could be extended  to all binaries in the WDS.  Indeed,
some of them contain known sub-systems in their secondaries.  However,
speckle interferometry  at SOAR  revealed many more  such sub-systems,
previously   missed  by   ``visual''  observers,   like  the   one  in
Figure~\ref{fig:speckle} \citep[see  also][]{THH10}.  They were missed
because  instruments used  normally  to observe  a  $1''$ binary  have
matched angular resolution  and do not allow discovery  of inner pairs
with separations much smaller than  $1''$.  Therefore, we do not apply
the  generic WDS  limit  (\ref{eq:WDS}) to  the secondary  components,
except the 8 historically resolved secondaries that do not have AO and
speckle coverage.

\subsection{Detection of spectroscopic binaries}
\label{sec:detrv}

Binary  companions  can  be  detected  from spectra  by  variable  RV,
appearance  of   double  lines,   or  presence  of   unusual  spectral
features.  The  RV variability  is  considered  here,  being the  most
general and powerful of those methods. 

The RV  data are  characterized by  the time span  $T$, the  number of
measurements $N_{\rm obs}$, and their intrinsic precision $\sigma_{\rm
  RV}$.  The generally accepted criterion of RV variability is related
to the normalized RV variance $\chi^2$, declaring all targets with the
low   $\chi^2$   probability  $P(\chi^2)   <0.01$   as  RV   variables
\citep{DM91,N04}.  This criterion  implies the false-alarm probability
of 1\%, meaning that about  40 false detections are expected among the
4080 FG-67  targets surveyed  by Nordstr\"om et  al.  Here we  use the
same $P(\chi^2)$  criterion, evaluate  the detection probability  as a
function  of   $(N_{\rm  obs},  T,  \sigma_{\rm   RV})$  by  numerical
simulation, and fit the results by a formula, as in \citep{Tok92}.

The semi-amplitude of  RV variation $K_1$ depends on  the period, mass
ratio, orbital inclination $i$, and eccentricity $e$,
\begin{eqnarray}
K_1 & = &  A_0 \sin i (1 - e^2)^{-1/2}, \nonumber \\
A_0 & = &  213 P^{-1/3} M_2 (M_1 + M_2)^{-2/3} .
\label{eq:K1}
\end{eqnarray}
Here $A_0$ (the semi-amplitude for a circular orbit at $i=90^\circ$, in
km\,s$^{-1}$) is related to the orbital period $P$ (in days) and the component
masses $M_1$ and $M_2$ (in solar-mass units).

In the simulations, we  assume normally distributed measurement errors
with  rms  $\sigma_{\rm  RV}$.   For  each value  of  three  detection
parameters $(N_{\rm obs}, T/P,  \kappa=A_0/\sigma_{\rm RV})$, a set of
1000  artificial  binaries  is  created.   Each binary  has  a  random
eccentricity $e$ distributed in the  interval [0,1] as $f(e) = (\pi/2)
\;  \sin (\pi  e)$, random  inclination $i$  (uniform  distribution of
$\cos  i$),   random  argument  of  periastron,   and  random  orbital
phase.  The  $N_{\rm  obs}$  random moments  of  observations  are  uniformly
distributed; the  interval  $T$  between the  first  and the  last
measurement  equals the specified  period fraction  $T/P$.  Obviously,
for  integer $T/P$ values the first and the last observations occur
at the  same orbital phase, in  this case SB1 cannot  be detected with
$N_{\rm obs}=2$.

There are two distinct regimes of spectroscopic binary detection. When
more than half of the orbit  is covered ($T/P > 0.5$), the duration of
observations does not matter  (except the above-mentioned case $N_{\rm
  obs}=2$), and the detection  probability $p_{\rm det}$ is a function
of only  two parameters $(N_{\rm obs},  \kappa)$.  On the  other hand, when
the observations  cover only  a small fraction  of the period,  $T/P <
0.1$, the  observed RV  variation is essentially  a linear  trend.  In
this regime, the first and  the last observations have most weight for
the detection, which becomes almost independent of $N_{\rm obs}$.  The
RV  variation  is  proportional  to  $T$, so  the  minimum  detectable
amplitude $ A_0/\sigma_{\rm RV} \propto 1/T$.  Indeed, the simulations
show that the curves $p_{\rm  det}(\kappa)$ are identical in the cases $T/P
= 0.1$ and  $T/P= 0.01$ if the arguments are scaled  10 times.  In the
intermediate  situation $0.1 <  T/P <  0.5$, the  detection probability
depends on all three parameters.

The detection probability found from the simulations can be fitted by the formula 
\begin{equation}
p_{\rm det}(y) 
=  \frac{y^\alpha }{  y^\alpha + 2 +  y^{-2.5}}, \;\;\; 
y = \frac{A_0}{\sigma_{\rm RV} \kappa_0}.
\label{eq:Pdet}
\end{equation}
The parameter  $\kappa_0$ equals  normalized amplitude where  the detection
probability is 1/4, the  parameter $\alpha$ regulates the steepness of
the curve  (steeper for larger $\alpha$). 

\begin{figure}
\plotone{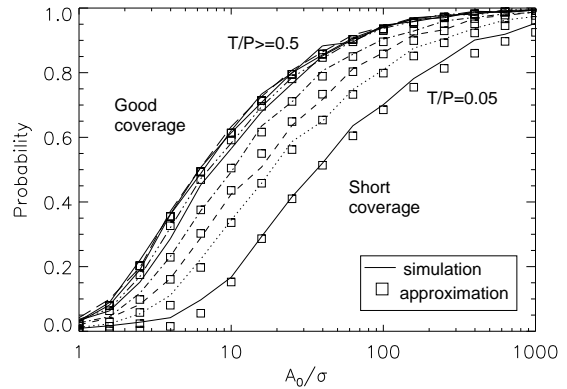}
\caption{Probability of detecting a spectroscopic binary $p_{\rm det}$
  as a function of $\kappa = A_0/\sigma_{\rm RV}$ for the case $N_{\rm obs}=3$ and
  increasing  period coverage  $T/P=(0.05, 0.1,  0.15, 0.2,  0.3, 0.5,
  0.7, 2.5, 13.5)$  (lines from right to left).   The analytical model
  is over-plotted as squares, the curves are results of simulation.
\label{fig:sb-det} }
\end{figure}

We describe the behavior of  the fitting parameters $(\kappa_0, \alpha)$ by
formulas that  work well  in both regimes and adequately represent
the transition (squares in Figure~\ref{fig:sb-det}):
\begin{eqnarray}
\kappa_0 & = & {\rm max} [ (1 + 0.65 P/T), 2.9 (N_{\rm obs} -1)^{-0.25} ]
\label{eq:x0} \\
\alpha & = & {\rm min} ( 0.7 + T/P, \alpha_0 ), 
\label{eq:alpha}
\end{eqnarray}
where $\alpha_0  = 0.95$ for $N_{\rm  obs} = 2$ and  $\alpha_0 = 1.25$
for $N_{\rm obs}  \ge 3$. In the short-coverage  regime, the values of
$(\kappa_0, \alpha)$ are defined by  the first terms and do not depend
on the  number of observations.  In  the good-coverage case  of $T/P >
0.5$, the  dependence on  $T$ vanishes, the  steepness of  the $p_{\rm
  det}(\kappa)$  curves  becomes  constant,  and  the  detection  threshold
$\kappa_0$ improves slowly with increasing number of observations
$N_{\rm       obs}$.         The       SB1       detection       model
(equations~\ref{eq:Pdet}--\ref{eq:alpha}) is not  very accurate (a few
percent  error),  but is  adequate  for  the statistical  description.
Compared  to the  imaging,  the RV  method  is more  ``probabilistic''
because $p_{\rm det}<1$ for a wide range of parameters.

Subsystems  in the  secondary  components of  visual  binaries can  be
detected by RV observations of the combined (blended) light if $\rho <
1\farcs5$ (e.g.  HIP~518 in  Figure~\ref{fig:dat}), but with a reduced
probability.  Many visual binaries in the FG-67 sample have periods $P
<100$\,yr, so that small  RV variations or trends are  attributable to the
motion of the visual binary and/or variable component blending, rather
than to a sub-system. We assume  that the RV detection of a sub-system
in the  blended spectra  is possible only  when the coverage  is good,
$T/P >2.5$,  and the  rms RV  of the blended  spectrum is  larger than
3\,km\,s$^{-1}$.   Alternatively,   moving  lines  of   the  binary  secondary
component B can be detected directly in the blended spectrum when they
are strong, $\Delta m_{AB} <1.5$, and well separated from the lines of
A, $A_0 > 15$\,km\,s$^{-1}$.

Both criteria  for detecting spectroscopic  binary in the  component B
were included  in the  simulations.  It turns  out that  the resulting
detection probability  can be described  by the equation~\ref{eq:Pdet}
with  parameters $y =  A_0(1-r)/[\kappa_0(1+r)]$, $\kappa_0  = (8  + 2
\Delta m_{AB})$,  and $\alpha =  2.5$, provided that $N_{\rm  obs} \ge
3$.  The additional factor $(1-r)/(1+r)$,  where $r = q^{3.75}$ is the
light  ratio in  the sub-system  Ba,Bb, accounts  for blending  of the
secondary lines;  when $r=1$, the blended  RV does not  change at all.
The  existing RV data  on close  visual binaries  are thus  useful for
constraining  the  frequency  of  spectroscopic  sub-systems  in  both
components,  although for the  secondary the  detection power  is much
less than for the primary.

A  few double-lined  binaries were  detected from  single  spectra. We
assign fake  detection parameters $T=100$\,d,  $N=3$, and $\sigma_{\rm
  RV}   =  2$\,km\,s$^{-1}$   to  cover   these  cases   by  the   same  model
(\ref{eq:Pdet}), but do not apply this recipe to single stars with one
RV datum.  A single RV measurement of both components of a wide binary
tells us that the binary is  physical if the RVs match.  In such case,
inner sub-systems  are unlikely, but this information  is not included
here in the detection model.

\subsection{Detection of astrometric binaries}
\label{sec:detacc}

For  the reasons  outlined  above,  we accept  only  the $\Delta  \mu$
binaries  from  \citet{MK05} as  valid  detections.   It  is shown  by
\citet{astrom1} that  orbital periods  of $\Delta \mu$  binaries range
from a few to a  hundred years.

The motion of  the photo-center $\mu_0$ (mas\,yr$^{-1}$) caused  by a ``dark''
companion in  a circular  face-on orbit is  related to  the semi-major
axis of  the astrometric orbit  $\alpha$ which, in turn,  is expressed
through the orbital period $P$ (years), primary mass $M_1$ (solar mass),
parallax $p_{\rm HIP}$ , mass ratio $q$, and light ratio $r$:
\begin{equation}
\mu_0 = \frac{ 2 \pi \alpha}{ P} =  
2 \pi P^{-1/3} M_1^{1/3} p_{\rm HIP} \frac{q -r}{(1 + q)^{2/3}(1 + r)} .
\label{eq:mu0}
\end{equation}
We assume $r = q^{3.75}$,  as appropriate for dwarfs less massive than
the Sun in the $V$ band.  

In the  simulations of  acceleration binaries, the  orbit orientation,
phase,  and  eccentricity  are  random (see  \S~\ref{sec:detrv}).  The
``orbital''  component of the  PM is  calculated by  linear fit  to 10
positions of the photo-center over  the time base of 3.2\,yr (duration
of the  {\em Hipparcos} experiment). Similar calculation  is done over
the 100\,yr time base of {\em Tycho}, and the difference gives $\Delta
\mu$.  To remove the  dependence on  parallax and  $q$, the  result is
normalized    by  $\mu_0$;  some  dependence on  the  binary  period
remains.

The cumulative distribution of  $\Delta \mu/\mu_0$ is approximated
analytically as
\begin{equation}
F(\Delta \mu/\mu_0 < \xi ) \approx y^a/(y^a + y^{-b} -1), \;\;\; y = (\xi/\xi_0).
\label{eq:dmu-model}
\end{equation}
The cumulative probability $F$ should be truncated at one (the formula
gives values $F>1$). 

The  parameters  of  the  approximating  formula  (\ref{eq:dmu-model})
$\xi_0$,  $a$, and  $b$  were  fitted to  the  simulations for  binary
periods  randing  from  3\,yr  to  500\,yr. Then  each  parameter  was
approximted  by a  polynomial of  $x  = \log  P$, enabling  analytical
calculation of $F$.

%with  $\xi_0  =  1.80  -  0.4/\log_{10}(P/1{\rm  yr})$,  $a=2.0$,  and
%$b=0.15$.

This statistical model allows us to evaluate the detection probability
for $\Delta \mu$ binaries.  For  each combination of the binary period
$P$ and  mass ratio $q$   -- a point in  the $(P,q)$ plane, --  we calculate
$\mu_0$  (the  parallax  is  known), assume  the  detection  threshold
$\Delta  \mu >  5$\,mas\,yr$^{-1}$  as appropriate  for  {\em Hipparcos},  and
calculate   $F(\Delta  \mu   <5)$  using   (\ref{eq:dmu-model}).   The
detection probability is $1-F$.  It peaks around $q \approx 0.5$.

The above model does not  account for binaries with short periods that
can  be  detected by  the  {\em  Hipparcos} acceleration  $\dot{\mu}$,
rather than by the $\Delta  \mu$ method. Most such binaries are either
confirmed spectroscopically (hence have  valid RV detection limits) or
rejected. A  few binaries with astrometric  orbits and no  RV data are
described by a fake RV coverage in the DET table.

\subsection{Detection of wide companions}

In this  sub-section, we cover  the detection of companions  with $\rho
>3''$ --  a  regime where    confusion  with  background  sources
increases with separation.

Companion  detection limits  in  the 2MASS  Point  Source Catalog  are
determined empirically  by plotting $\Delta  K_s (\rho)$ \citep[Fig.~8
  in ][]{ANDICAM}.  We adopt the ``realistic'' limit for $3'' < \rho <
30''$, see Table~\ref{tab:lim}. In crowded fields, companions selected
from  2MASS by  their colors  are confused  with background  stars. We
characterize crowding by  $N^*$ -- the number of  2MASS sources within
$150''$ from  the target.   It is assumed  here that 730  targets with
$N^* > 100$  are not screened for companions  with 2MASS (however, the
{\em  Hipparcos}   and  WDS  detection  limits  are   still  valid for
them).

Wide binaries  are also found in the  WDS. The large time  base of the
WDS  allows discrimination  of  optical companions  by their  relative
motion, hence  the WDS threshold in Table~\ref{tab:lim}  is applied to
all targets out to $30''$ separation.

At  separations $\rho  > 30''$,  we  used the  SUPERBLINK catalog  and
selected  CPM  companions  by  both  color and  PM  \citep{LEP}.   The
companion search is 90\% complete to $V=19^m$.  Only 2966 targets
(61\% of  the sample) with  PM above the SUPERBLINK  limit (40\,mas\,yr$^{-1}$
north of $-20^\circ$ and 150\,mas\,yr$^{-1}$ otherwise) are covered.  Most CPM
companions have separations $\rho  < 300''$ (projected separation less
than 20\,000\,AU at 67\,pc).

CPM pairs wider  than $30''$ are also found in the  WDS or by matching
the {\em  Hipparcos} stars in PM  and parallax \citep[e.g.][]{SO2011}.
Some  of these binaries  have PM  below the  SUPERBLINK limit  or were
missed by it for other reasons.  We add fictitious imaging data to the
DET table to cover those exceptions, with a reference 'CPM'.

Note that  identification of wide  binaries by common PM  and matching
colors  introduces a bias against  hierarchical multiples  because their
sub-systems  perturb  both  PM  and  photometry.   Partially  resolved
secondary  components  will  not   appear  in  the  catalogs  such  as
SUPERBLINK.  As this  bias is  difficult to  quantify, it  is silently
ignored in the  statistical analysis.

\subsection{Average detection probability}

\begin{figure}
\plotone{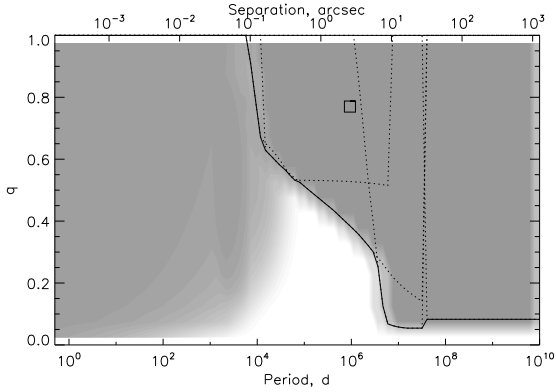}
\caption{Probability  of  companion   detection  for  HIP~55  ($p_{\rm
    HIP}=15.4$\,mas). Individual  limits for resolved  companions from
  {\em Hipparcos},  WDS, 2MASS, and  CPM are plotted in  dotted lines,
  the combined  limit --  in full  line.  No data  from AO  or speckle
  interferometry are available for  this target.  The visual companion
  B at  3\farcs8 (square) is  listed in the  WDS and 2MASS.   The gray
  shading  shows the  combination  of spatial  resolution with  limits
  resulting from  {\em Hipparcos}  acceleration $\Delta \mu$  and from
  the  five   RV  measurements  over  time   interval  $T=1394$\,d  by
  \citet{N04}.
\label{fig:HIP55} }
\end{figure}

\begin{figure}
\epsscale{1.15}
\plotone{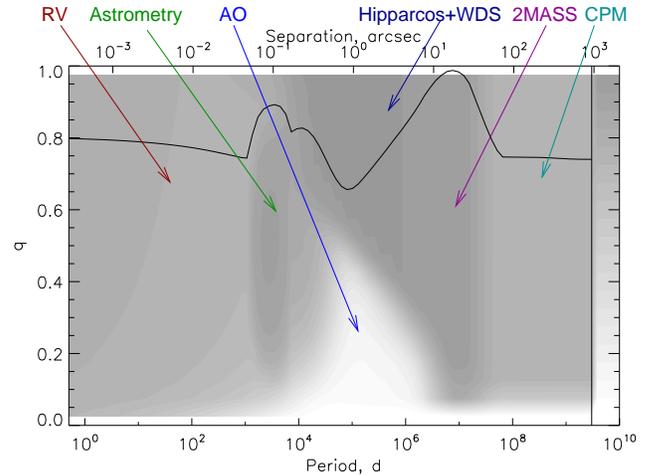}
\caption{Average  probability of  companion detections  around primary
  targets as function of period $P$ and mass ratio $q$.  The color bar
  on the  right indicates the  scale, from $p_{\rm det}=0$  (white) to
  $p_{\rm det}=1$ (gray). The  black curve shows detection probability
  averaged over $q>0.1$.   Dominant detection techniques are indicated
  on the top.  The upper  scale shows angular separation at a distance
  of 50\,pc.
\label{fig:det} }
\end{figure}

The  probability   of  companion  detection  is   evaluated  for  each
individual  primary  target.  First,  the  resolution  limits of  {\em
  Hipparcos} and WDS are applied to all targets. The limits from 2MASS
are added  for targets with $N^*  < 100$. The  SUPERBLINK limit $V<19$
and $\rho>30''$ is applied to targets with a fast enough PM.  Then the
individual  imaging  limits  are   added  from  the  DET  table,  when
available.  The  resulting curve  $q_{\rm min}(P)$ splits  the $(P,q)$
plane in two parts, with $p_{\rm det} = 1$ above it and $p_{\rm det} =
0$ below.  This  sharp limit is softened to account  for the fact that
the detection  depends on  the apparent (projected)  separation which,
for a  given period,  is random (see  \S~\ref{sec:syspar}).  If,  at a
given  mass ratio,  the companion  becomes resolvable  at  some period
$P^*$, we assign detection probability  of 1/3 to periods $P^*/1.6 < P
< P^*$, 2/3 to $P^* < P < 1.6 P^*$, and one to longer
periods.

\begin{figure}
\epsscale{1.1}
\plotone{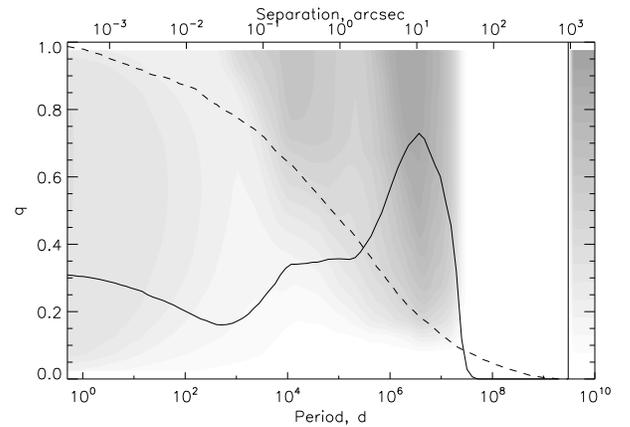}
\caption{Average probability of detecting sub-systems in the secondary
  components. The  color scale is from  0 (white) to  1 (gray), the
  black curve shows detection probability averaged over $q$ and over
  relevant secondaries, the dashed line shows the fraction of relevant secondaries.
\label{fig:detsec} }
\end{figure}

The  probability  of  resolving  a  companion  is  combined  with  the
astrometric  detection (equation~\ref{eq:dmu-model})  and with  the RV
detection (if  the RV  data are available)  to evaluate  the detection
probability  $p_{\rm det} =  1 -  \Pi_i (1  -p_i)$ resulting  from the
combination of  $i$ independent techniques.   Calculation of detection
limit for one target is illustrated in Figure~\ref{fig:HIP55}.

Figure~\ref{fig:det} shows
the average $p_{\rm  det} (P,q)$ for all primary  targets in the FG-67
sample.  The combination of  observing techniques covers the parameter
space rather uniformly, with the exception of low-mass companions with
separations from  $0.3''$ to $10''$ that  are too close  for the 2MASS
and  too faint  for the  WDS. Such  companions can  be probed  with AO
(which, so far, was used only on a small subset of targets).

For  the  {\em  secondary}  components the  detection  probability  is
evaluated  separately.   The   {\em  Hipparcos}  detection  limits  in
resolution  and  acceleration apply  to  the  bright secondaries  with
individual HIP numbers and separations $>20''$ (at closer separations,
the two  stars were  treated by  {\em Hipparcos} as  a binary  and any
sub-systems  were  missed  because  resolved  triple  stars  were  not
considered in the data  reduction).  The secondary components observed
individually in RV  or by imaging are also treated in  the same way as
the primaries. Additionally, detection  data on the primary components
of  close  visual  binaries  are  used  to  constrain  sub-systems  in
their secondaries, as explained in \S\ref{sec:detvis} and \S\ref{sec:detrv}.

Figure~\ref{fig:detsec}   illustrates  the   average   probability  of
detecting sub-systems  in the  secondary components.  The  curve shows
the  probability averaged  over  $q >  0.05/M_2$  (i.e.  ignoring  any
brown-dwarf companions)  for the relevant secondaries  (at each period
of a  potential sub-system, relevant secondaries are  those where such
subsystems  are  allowed  dynamically  by the  outer  binaries).   The
proportion of  relevant secondaries is  plotted as dashed  curve. Only
1676 secondaries with estimated mass  (77\% of their total number) are
considered.  There are no  constraints on sub-systems in the remaining
23\%  of secondaries  belonging to  the spectroscopic  and astrometric
binaries with unknown periods.

The estimated detection rate of sub-systems in the secondary components is
less  than for  the primaries,  about 0.3  at short  periods,  0.16 at
periods around 1\,yr, and 0.34  at periods around 30\,yr.  The average
probability of detecting a  secondary sub-system with $P \sim 10^4$\,d
and  $q>0.8$  is about  0.58,  owing  to  the high-resolution  imaging
surveys of  solar-type stars with  RoboAO \citep{RoboAO} and  at SOAR.
At  separation $>3''$,  the sub-systems  are constrained  by the classical
imaging  (e.g.   2MASS).   A  targeted  RV survey  of  secondaries  is
obviously needed to reach better coverage at short periods.  More than
a  half of sub-systems  in the  secondaries still  remain undiscovered.
Previous  work  on  multiplicity  \citep[e.g.][]{R10} focused  on  the
companions  to main targets  and neglected  potential binarity  of the
secondaries  (some  binary  secondaries  within 25\,pc  are  recovered
here).

\begin{deluxetable*}{rrrc cc rr ccc c  cccc c} 
\tabletypesize{\scriptsize}      
\tablecaption{COMP: data on individual components
\label{tab:ncomp}   }     
\tablewidth{0pt} 
\tablehead{ HIP0  & HIP & HD & Comp. & $\rho$ & $\theta$ & \multicolumn{2}{c}{R.A., Dec. (J2000)} &
 $\mu_{\alpha *}$ & $\mu_\delta$ & $p_{\rm HIP}$ & \ldots &
$V$ &  $I_C$ & $J$ & $K_s$ & $N^*$  \\
% -- header line 2 
  &  &   &   & $''$ & deg & deg & deg &  \multicolumn{2}{c}{mas yr$^{-1}$} & mas & \ldots&
 mag & mag &  mag &  mag & \\
% header line 3
(1) & (2) & (3) & (4) &  (5) & (6)& (7) & (8) & (9) & (10) & (11) &\ldots& (15) & (16) & (17) & (18) & (19) }
\startdata
50    &50    &224782&A  &   0.0&   0.0&    0.143085&  $-$53.097713&   52.9&  $-$20.9&  16.8&\ldots&  6.49&  5.81&  5.41&  5.05& 14 \\
50    &0     &0     &B  &   1.6& 331.0&    0.142726&  $-$53.097325&    0.0&      0.0&  16.8&\ldots&  9.85&  0.00&  0.00&  0.00& 14 \\
55    &55    &224783&A  &   0.0&   0.0&    0.158834&  $-$66.683174&  162.8&  $-$28.9&  15.4&\ldots&  7.40&  6.78&  6.53&  6.21& 18 \\
55    &0     &0     &B  &   3.8& 274.0&    0.156173&  $-$66.683100&    0.0&      0.0&  15.4&\ldots&  9.17&  0.00&  6.80&  6.40& 18 \\
58    &58    &224792&A  &   0.0&   0.0&    0.173513&   62.175899&  $-$46.9&  $-$43.8&  25.8&\ldots&  7.05&  6.46&  6.07&  5.81&208 \\
81    &81    &224828&A  &   0.0&   0.0&    0.243409&  $-$4.932534&$-$184.6& $-$172.6&  22.8&\ldots&  8.57&  7.86&  7.34&  6.96& 10 
%93    &93    &224839&A  &   0.0&   0.0&    0.292358&   $-$0.076098&   54.9&  $-$73.2&  16.3&\ldots&  8.12&  7.43&  7.00&  6.67& 14 \\
%135   &135   &224908&A  &   0.0&   0.0&    0.426798&   $-$0.217777&   20.7& $-$114.0&  20.0&\ldots&  8.64&  7.95&  7.51&  7.17&  6 \\
%142   &142   &224918&A  &   0.0&   0.0&    0.453783&        66.306033&    4.7&  12.3&  16.8&\ldots&  7.33&  6.55&  6.07&  5.75&114
\enddata                                                                                                                       
\tablenote{Table  4 is  published in  its entirety  in  the electronic
 edition of  AJ, a portion is  shown here for  guidance regarding its  form and content.}
\end{deluxetable*}

% Systems: sys.txt produced by expsys in sys.pro
\begin{deluxetable*}{r ll rcr cc rc rc rc l}
\tabletypesize{\scriptsize}      
\tablecaption{SYS: data on binary systems 
\label{tab:sys}   }     
\tablewidth{0pt} 
\tablehead{ HIP0  & Comp. & Type & \multicolumn{2}{c}{$\rho$} & $\theta$ & $V_1$ & $V_2$ &  \multicolumn{2}{c}{$P$} & 
\multicolumn{2}{c}{$M_1$} & \multicolumn{2}{c}{$M_2$} & Comment \\
(1) & (2) & (3) &   \multicolumn{2}{c}{(4-5)} & (6)& (7) & (8) & \multicolumn{2}{c}{(9-10)} & \multicolumn{2}{c}{(11-12)} & \multicolumn{2}{c}{(13-14)} & (15) }
% -- header line 2 
%  &  &   &   & $''$ & deg & deg & deg &  \multicolumn{2}{c}{mas yr$^{-1}$} & mas & \ldots&
% mag & mag &  mag &  mag & \\
% header line 3
\startdata
    50&A,B,*     &Ch      &  1.49&"&331.0& 6.55& 9.85& 534.99&y&1.55&v&0.87&v&HJ\_5437                   \\
    55&A,B,*     &Chp     &  3.800&"&274.0& 7.69& 9.17&   2.556&k&1.31&v&1.00&v&GLI\_289                   \\
    93&Aa,Ab,*   &s,a,v   &  0.330&"& 73.0& 8.12& 15.98& 76.17&y&1.18&v&0.24&k&CfA:P=long MK05:dmu NICI     \\
   179&Aa,Ab,*   &S2       &  2.008&m&  0.0& 6.91& 0.00& 10.674&d&1.32&v&1.32&q&N04:drv=30.5* Orb.Gorynya2013 \\
   223&A,B,*     &v       &  1.600&"&163.0& 7.40& 9.36& 397.08&y&1.17&v&0.84&v&BU\_281AB                  \\
   223&X A,C     &Xph     & 44.300&"&330.0& 7.40&11.70&   0.000&y&0.00&-&0.00&-&HJ\_998AC Reflex PM          \\
   290&Aa,Ab,*   &s,a     &  0.000&"&  0.0& 7.78& 0.00&   0.000&y&1.29&v&0.00&-&N04:drv=1.3 MK05:dmu       \\
   305&X Aa,Ab   &a?      &  0.000&"&  0.0& 7.81& 0.00&   0.000&y&1.13&v&0.00&-&HIP:7 N04:RV=const WSI:UR 
\enddata                                                                  
\tablenote{Table  5 is  published in  its entirety  in  the electronic
 edition of  AJ, a portion is  shown here for  guidance regarding its  form and content.}
\end{deluxetable*}                                                                                              

\begin{deluxetable}{r l c c c l  }
\tabletypesize{\scriptsize}      
\tablecaption{DET1: detection limits in radial velocity
\label{tab:hipdet1}   }     
\tablewidth{0pt} 
\tablehead{ HIP0  & Comp & $T$ &  $N_{\rm obs}$  & $\sigma_{\rm RV}$ & Ref. \\
                  &      & d   &               & km\,s$^{-1}$ & } 
\startdata
    50 & AB &    683 &    3 &    0.300 & N04       \\
    50 & B  &    208 &    3 &    0.020 & Nid02     \\
    55 & AB &   1394 &    5 &    0.400 & N04       \\
    58 & A  &   1168 &    2 &    0.300 & N04       \\
    58 & A  &   4109 &    9 &    0.300 & CfA       \\
    81 & A  &   4433 &   13 &    0.610 & Latham2002   
\enddata                                                                
\tablenote{Table  6 is  published in  its entirety  in  the electronic
 edition of  AJ, a portion is  shown here for  guidance regarding its  form and content.}
\end{deluxetable}

\begin{deluxetable*}{l cc cccc  cccc l} 
\tabletypesize{\scriptsize}      
\tablecaption{DET2: imaging detection limits
\label{tab:hipdet2}   }     
\tablewidth{0pt} 
\tablehead{HIP0 & Comp & $\lambda $ & $\rho_1$& $\rho_2$& $\rho_3$&$\rho_4$&  
$\Delta m_1$&  $\Delta m_2$ &$\Delta m_3$ &$\Delta m_4$& Ref. \\
(1) & (2) & (3) & (4) &  (5) & (6)& (7) & (8) & (9) & (10) & (11) & (12)  }
\startdata
    50 &A &  540.0 & 0.029 & 0.150 & 1.000 &  1.500 & 0.50 & 5.51 & 6.50 & 6.50& SOAR \\        
    81 &A &  550.0 & 0.030 & 0.100 & 0.200 &  1.500 & 0.00 & 1.35 & 2.70 & 3.00& WSI  \\        
    93 &A & 2272.0 & 0.054 & 0.150 & 0.900 &  9.000 & 0.00 & 5.26 & 7.15 & 7.15& NICI \\        
   135 &A &  550.0 & 0.030 & 0.100 & 0.200 &  1.500 & 0.00 & 1.35 & 2.70 & 3.00& WSI  \\        
   179 &A &  550.0 & 0.066 & 0.220 & 0.439 &  1.500 & 0.00 & 0.35 & 2.70 & 3.00& INT4 \\        
   179 &A &  770.0 & 0.150 & 0.800 & 2.100 & 13.920 & 2.65 & 5.19 & 6.29 & 6.29& RoboAO 
\enddata                                                                                                                       
\tablenote{Table  7 is  published in  its entirety  in  the electronic
 edition of  AJ, a portion is  shown here for  guidance regarding its  form and content.}
\end{deluxetable*}

%\setcounter{table}{7}

% Notes on individual systems: notes2.txt produced by notes.pro ==> expnotes2
\begin{deluxetable*}{r l}
\tabletypesize{\scriptsize}      
\tablecaption{Notes 
%\tablenumber{9}
\label{tab:notes}   }     
\tablewidth{0pt} 
\tablehead{ HIP0  & Text }
\startdata
    93& CfA: P=long N04:drv=1.2. SIMBAD: 14 ref. \\
    93& Tok2012bis: Resolved with NICI at 0.32". 72d, dK=4.3 dH=4.3. V(AB) is estimated \\
   179& Speckle: 2001AJ....121.3224M  SIMBAD: 12 ref. N04:q=1.00+-0.011 N=5 drv=30.5 \\
   179& SB2 orbit by Gorynya2013: P=10.674d, e=0.34,  K1=55.00 K2=56.62, twin. \\
   276& Cfa: SB, P=538d? 
\enddata                                                                
\tablenote{Table  8 is  published in  its entirety  in  the electronic
 edition of  AJ, a portion is  shown here for  guidance regarding its  form and content.}
\end{deluxetable*}

%-------------------------------------------------------------
\section{Description of the tables}
\label{sec:tables}

%-------------------------------------------------------------
\subsection{COMP: data on individual components}
\label{sec:main}

Table~\ref{tab:ncomp}    contains   identifiers,    coordinates,   and
photometry  of known  components,  both primary  and secondary.   Most
resolved  secondary  components  with  separations  above  $1''$  have
individual entries in  Table~\ref{tab:ncomp}.  Some bright secondaries
have  HIP or  HD numbers.   The  notion of  ``resolved component''  is
fuzzy,  however, so a  few components  are not  in this  Table despite
having  $\rho >1''$.  Optical  components are  not included,  except a
few.

The columns of Table~\ref{tab:ncomp}  contain (1) HIP0 number, (2) HIP
number which  equals HIP0 for  the primary target, (3)  HD identifier,
and (4) component designation  by a capital letter, generally matching
the WDS designations.  Then follow  (4) separation from the primary in
arcseconds and (5) position angle of the companion.  Both these fields
are zero for primary targets.  Columns (9) and (10) contain equatorial
coordinates  for J2000 in  the ICRS  system, taken  from HIP2  in most
cases. If  only the  relative position of  the secondary  component is
known from the  WDS, its coordinates are calculated  with reference to
the primary.  Coordinates of  some secondary components are taken from
2MASS.

Columns (9), (10), and (11) list the PM $\mu_{\alpha}^*$, $\mu_\delta$
and the  parallax $p_{\rm HIP}$ taken  from HIP2 for  the main targets
and  for  the  secondaries  with  separate  HIP  entries.   For  other
secondary components, the  parallax equals that of the  main target by
definition, the PM is zero,  if not measured independently. Sources of
PM  for the  secondary components  are WDS,  PPMX  \citep{PPMX}, NOMAD
\citep{NOMAD},  or SUPERBLINK.   Errors  of the  PM  and parallax  are
listed  in  columns  (12-14)  when available,  zero  otherwise  (these
columns are  omitted from  the printed fragment  of the table  to save
space).

Magnitudes  in the  $V$,  $I_C$, $J$,  and  $K_s$ bands  are given  in
columns  (15-18),  respectively, taken  mostly  from  HIP2 and  2MASS.
Other  sources (e.g.  WDS and  NOMAD)  are invoked  to complement  the
photometry  of secondary  components,  whenever possible.  Photometric
information helps to identify physical companions by their position on
the CMDs.  The last  column (19) of Table~\ref{tab:ncomp} lists $N^*$,
the number  of point  sources in the  2MASS catalog within  $150''$ of
each  primary   target,  to  quantify  the  density   of  the  stellar
background.

%-------------------------------------------------------------
\subsection{SYS: data on binary systems}
\label{sec:sys}

A  hierarchical multiple  system consists  of several  binaries, where
some components  are actually pairs  of stars.  Each binary  system or
sub-system corresponds to a  line in Table~\ref{tab:sys}.  It contains
the {\it  Hipparcos} number of the  primary target HIP0  in column (1)
and the  designation of  the system by  a comma-separated list  of its
components (primary, secondary, parent) in column (2).  Then in column
(3)  the type  of the  system is  listed by  codes explained  above in
\S~\ref{sec:data}.   Optical pairs  from  the WDS  and other  spurious
binaries  are  kept  in  the   SYS  table  for  completeness  and  are
distinguished by their component designation starting with 'X'.

The following columns  (4, 5, 6) list the separation  and its units ("
for  arcseconds, m  for  milliarcseconds) and  the  position angle  in
degrees (zero  if unknown).  Visual  magnitudes of the  components are
given in the  columns (7, 8).  The orbital period  and its units ('d',
'y', 'k' for days, years,  and kiloyears, respectively) are in columns
(9, 10).  The estimated masses of the primary and secondary components
with 1-letter codes indicating the method (see \S\ref{sec:syspar}) are
listed  in columns  (11, 12)  and  (13, 14),  respectively.  The  last
column (15)  of Table~\ref{tab:sys} contains a  short comment pointing
to  the source  of the  information  (e.g.  WDS  discoverer codes  for
resolved binaries, SB9 references  for SBs).  Bibliographic codes from
Table~\ref{tab:bib} are used extensively in the comments.

\subsection{DET: individual detection limits}

Methods  used  to  estimate   the  detection  limits  are  covered  in
\S\ref{sec:det}.  The  data on individual components  are presented in
two tables, DET1 and DET2.  Table~\ref{tab:hipdet1} (7109 lines) lists
the RV data, linked to the  particular component by the HIP0 number in
column (1) and the component  identifier in column (2).  Then the time
coverage $T$  in days, the  number of observations $N_{\rm  obs}$, and
the measurement precision $\sigma_{\rm  RV}$ in km\,s$^{-1}$ are given
in  columns  (3--5),   respectively,  followed  by  the  bibliographic
reference in column (6).  The  reference codes match the references in
Table~\ref{tab:bib}.

The  detection  limits of  imaging  (AO  and  speckle) are  listed  in
Table~\ref{tab:hipdet2} (DET2,  4165 lines).  The columns  (1) and (2)
contain the  HIP0 and component, as  in Table~\ref{tab:hipdet1}.  Then
in column (3)  the imaging wavelength in nm  is given.  Columns (4--7)
list the  separations $\rho_i$ in arcseconds,  columns (8--11) contain
the corresponding detection limits $\Delta m_i$, and column (12) gives
the reference code.

\subsection{Notes}

Notes are  given in Table~\ref{tab:notes}  as free text linked  to the
HIP0 number of  the target.  The references are  denoted by the codes
from Table~\ref{tab:bib} or given  explicitly in the
notes.

%-------------------------------------------------------------
\section{Overview}
\label{sec:overview}

The SYS  table contains  3068 pairs, 2196  of which are  physical, the
rest are optical  or spurious. Periods of 357  binaries remain unknown
(208 of 'a' type  and 261 of 's' type, with an  overlap of 112 between
those  groups). It  is safe  to assume  that all  unknown  periods are
shorter  than 100\,yr. The  proportion of  unknown periods  among 1132
systems with  $P<100$\,yr is 32\%.

This rich material  is used in the accompanying  Paper~II to study the
statistics of  hierarchical stellar systems.  The  data collected here
can be  useful for several  other purposes, for example  to complement
exo-planet  programs, to  search for  Sirius-like binaries,  to select
fast resolved pairs for orbit calculation, or to study relative motion
in wide binaries and resolved triples.  The weakness of this sample --
missing information on many  spectroscopic and astrometric binaries --
can be  corrected in the  future by RV monitoring  and high-resolution
imaging.

\acknowledgments I am grateful  to D.~Latham (CfA) and B.~Mason (USNO)
for sharing  their unpublished data.  This project  benefited from 
fruitful   collaboration  with  M.~Hartung,   S.~L\'epine,  R.~Riddle,
N.~Gorynya, and others.

This work  used the  SIMBAD service operated  by Centre  des Donn\'ees
Stellaires  (Strasbourg, France),  bibliographic  references from  the
Astrophysics Data System maintained  by SAO/NASA, data products of the
Two Micron All-Sky Survey  (2MASS), the Washington Double Star Catalog
maintained at USNO, and the  SB9 catalog managed by D.~Pourbaix. It is
a suitable occasion  to celebrate the often neglected  effort of those
who  maintain  catalogs  and  databases  and  thus  provide  keep  the
foundation of astronomy solid.

{\it Facilities:}  \facility{SOAR}, \facility{SMARTS}

%\clearpage


\begin{thebibliography}{}


\bibitem [Abt \& Willmarth (2006)]{Abt2006}
Abt, H. A. \& Willmarth, D. 2006, ApJS, 162, 207

\bibitem[Anderson \& Francis (2012)]{XHIP}
Anderson, E. \& Francis, C. 2012, AstL, 38, 331

\bibitem[Baraffe et al. (1998)]{Baraffe}
Baraffe, I., Chabrier, G., Allard, F., \& Hauschildt, P. H.  1998, A\&A,
337, 403


\bibitem[Bessell \& Brett (1988)]{Bessell}
Bessell, M. H., \& Brett, J. M. 1988, PASP, 1134 

\bibitem[Caballero (2010)]{Caballero2010}
Caballero, J. A. 2010, A\&A, 514, 98

\bibitem[Casagrande et al. (2013)]{Casagrande}
Casagrande, L., Sch\"onrich,  R.,  Asplund, M. et al. 2013, A\&A, 530, 138

\bibitem[Chauvin et al. (2006)]{Chauvin06}
Chauvin, G., Lagrange, A.-M., Udry, S. et al. 2006, A\&A, 456, 1165
% Fusco, T., Galland, F., Naef, D., Beuzit, J.-L., Mayor, M.2006A&A...456.1165C 

\bibitem[Chauvin et al. (2010)]{Chauvin10}
Chauvin, G., Lagrange, A.-M., Bonavita, M. et al. 2010, A\&A, 509, 52

\bibitem[Cutri et al. (2003)]{2MASS}
Cutri,  R.  M.,  Skrutskie, M.  F.,  van  Dyk,  S. et al. 2003
%Beichman,  C.  A., Carpenter, J.  M., Chester, T.,  Cambresy, L., Evans, T.,  Fowler, J.,
%Gizis,  J.  et  al.   2003,   
The  IRSA  2MASS  All-Sky  Point Source  Catalog.  NASA/IPAC  Infrared
Science Archive.

\bibitem[Delfosse et al. (2000)]{Delfosse}
Delfosse, X., Forveille, T., S\'egransan, D. et al. 2000, A\&A, 364, 217


\bibitem[Duquennoy \& Mayor (1991)]{DM91}
Duquennoy, A. \& Mayor, M. 1991, A\&A, 248, 485

\bibitem[Eggenberger et al. (2007)]{Egg2007}
Eggenberger, A., Udry, S., Chauvin, G. et al. 2007, A\&A, 474, 273

\bibitem[Fabrycky \& Tremaine (2007)]{Fabrycky2007}
Fabrycky, D. \& Tremaine, S. 2007,  ApJ, 669, 1298

\bibitem[Frankowski et al. (2007)]{Frankowski}
Frankowski, A., Jancart, S., \& Jorissen, A. 2007, A\&A, 464, 377

\bibitem[Ginski et al. (2012)]{Ginski2012}
 Ginski, C., Mugrauer, M.,  Seeliger, M., \& Eisenbeiss, T. 2012, MNRAS, 421, 2498

\bibitem[Girardi et al. (2000)]{Girardi}
Girardi, L., Bressan, A., Bertelli, G., \& Chiosi, C., 2000, A\&A,
141, 371

\bibitem[Goldin \& Makarov (2006)]{GM06}
Goldin, A. \& Makarov, V. V. 2006, ApJS, 166, 341 


\bibitem[Griffin (2012)]{Griffin2012}
Griffin, R. F. 2012, JAA, 33, 29
% Hyades spectroscopic binaries. Papers/Griffin2012.pdf


\bibitem[Halbwachs et al. (2012)]{Halb2012}
Halbwachs, J.-L., Mayor, M., \& Udry, S. 2012, MNRAS, 422, 14

\bibitem[Hartkopf et al. (2001)]{INT4}
 Hartkopf, W. I., Mason, B. D., \& McAlister, H. A.   2001, AJ, 122, 3480
\url{ http://ad.usno.navy.mil/wds/int4.html}

\bibitem[Hartkopf \& Mason (2013)]{VB6}
Hartkopf, W. I. \& Mason, B. D. 2013,
Sixth Catalog of Orbits of Visual Binary Stars.
USNO \url{http://ad.usno.navy.mil/wds/orb6.html}

\bibitem[Hartkopf et al. (2013)]{Hartkopf13}
Hartkopf, W. I., Mason, B. D., Finch, C. T. et al.
2013, AJ, 146, 76


\bibitem[Henri \& McCarty (1993)]{HM93}
Henry, T. J. \& McCarthy, D. W. 1993, AJ, 106, 773

\bibitem[Holberg et al. (2013)]{Holberg13}
Holberg, J. B., Oswalt, T. D., \&  Sion, E. M.
2013, MNRAS, 435, 2077

\bibitem [Horch et al. (2011)] {Horch2011}
Horch, E. P., Gomez, S. C., Sherry, W. H. et al. 2011, AJ,  141, 45


\bibitem[Janson et al. (2013)]{SEEDS}
Janson, M., Brandt, T. D., Moro-Mart\'{i}n, A. et al. 
2013, ApJ, 773, 73

\bibitem[Jenkins et al. (2010)]{Jenkins2010}
Jenkins, J. S., Jones, H. R. A., Biller, B. et al. 2010, A\&A, 515, 17

\bibitem [J\'odar et al. (2013)]{Jodar2013}
J\'odar, E., P\'erez-Garrillo, A., D\'{i}az-S\'anchez, A. et al.
2013, MNRAS, 429, 859

\bibitem[Jones et al.  (2002)]{Jones2002}
Jones, H. R. A., Butler P. R., Marcy, G. W. et al. 2002, MNRAS, 337, 1170


\bibitem[Lafreni\`ere et al. (2007)]{LAF07}
Lafreni\`ere, D., Doyon, R., Marois, Ch. et al. 2007, ApJ, 670, 1367

\bibitem[Lagrange et al. (2009)]{Lagrange2009}
Lagrange, A.-M., Desort, M., Galland, F. et al. 2009,  A\&A, 495, 335

\bibitem[Lang (1992)]{Lang92}  
Lang, K.  R. 1992, Astrophysical  data. Planets
  and Stars (Berlin: Springer-Verlag)

\bibitem [Latham et al. (2002)] {Latham2002} Latham, D. W., Stefanik,
  R. P., Torres, G., \&  Davis, R. J.  2002 AJ, 124, 1144

\bibitem[Makarov \& Kaplan (2005)]{MK05}
Makarov, V. V. \& Kaplan, G. H.,  2005, AJ, 129, 2420

\bibitem[Mason et al.\ (2001)]{WDS}
Mason, B. D., Wycoff, G. L., Hartkopf, W. I., Douglass, G. G. \&
Worley, C. E. 2001, AJ, 122, 3466 
%(see the current version at 
\url{http://www.usno.navy.mil/USNO/astrometry/optical-IR-prod/wds/wds.html}

\bibitem[Metchev \& Hillenbrand (2009)]{MH09}
Metchev, S. A. \& Hillenbrand, L. A. 2009, ApJS, 181, 62


\bibitem[Nidever et al. (2002)]{Nid02}
Nidever, D. L., Marcy, G. W., Butler, R. P. et al. %, Fischer, D.A.,Vogt, S.S. 
2002, ApJS, 141, 503

\bibitem[Nordstr\"om et al. (2004)]{N04}
%Nordstr\"oem et al. 2004
Nordstr\"om,  B., Mayor,  M.,  Andersen, J.  et al.  2004, A\&A, 418, 989 
%Holmberg, J., Pont,  F., Jorgensen, B. R., Olsen, E.  H., Udry, S., \&
%Mowlavi, N.  2004, A\&A, 418, 989 (GCS)

\bibitem[Peters \& Fabrycky (2009)]{BlueStrugglers}
Perets, H. B. \& Fabrycky, D. C. 2009, ApJ, 134, 2353

%\bibitem[ESA (1997)]{HIP1}
\bibitem[Perryman \& ESA (1997)]{HIP1}
Perryman, M. A. C. \& ESA 1997, in ESA Publ. Ser. 1200,
The Hipparcos and Tycho Catalogues,
(Noordwijk: ESA), Vol. 1
%ESA 1997, The Hipparcos and Tycho Catalogues, ESA SP-1200

\bibitem[Pourbaix et al. (2004)]{SB9}
Pourbaix, D., Tokovinin, A. A., Batten, A. H. et al. 2004 A\&A, 424, 727
%Fekel F.C.,  Hartkopf W.I., Levato  H., Morrell N.I., Torres  G., Udry
%S. SB9: The Ninth Catalogue of Spectroscopic Binary Orbits. 2004, A&A, V. 424, PP. 727-732

\bibitem[Raghavan et al. (2010)]{R10}
Raghavan, D., McAlister, H. A., Henry, T. J. et al.  2010, ApJS, 190, 1
%Latham, D. W., Marcy, G. W., Mason,  B. D., Gies, D. R., White, R. J.,
%\& ten Brummelaar, Th. A. 2010, ApJS, 190, 1

\bibitem[Rameau et al. (2013)]{Rameau2013}
Rameau, J., Chauvin, G., \& Lagrange, A.-M. 2013, A\&A, 553, 60

\bibitem[Reed et al. (2014)]{RoboAO}
Reed, R., Tokovinin, A., Mason, D. B.  et al, 2014, AJ, in preparation.

\bibitem[Roell et al. (2012)]{Roell2012}
Roell, T., Neuh\"auser, R., Seifahrt, A. et al.   %, \& Mugrauer, M.
2012, A\&A, 92

\bibitem[Roeser et al. (2008)]{PPMX}
Roeser, S., Schilbach, E., Schwan, H. et al. 2008, A\&A, 488, 401

\bibitem[Santerne et al. (2013)]{Santerne2013}
Santerne, A., Fressin, F., D\'{i}az,  R. F. et al. 2013, A\&A, 557, 139

\bibitem[Shaya \& Olling (2011)]{SO2011}
Shaya, E. J. \& Olling, R. P. 2011, ApJS, 192, 2

\bibitem[Soubiran el al. (2010)]{PASTEL}
Soubiran, C., Le Campion, J.-F., Cayrel de Strobel, G.,
\& Caillo, A. 2010, A\&A, 515, 111


\bibitem[Tokovinin (1992)]{Tok92}
Tokovinin, A. A. 1992, A\&A, 256, 121


\bibitem[Tokovinin (1997)]{MSC}
Tokovinin, A. 1997, A\&AS 124, 75 
%(see the curent version at 
\url{http://www.ctio.noao.edu/\~{}atokovin/stars/index.php}

% \bibitem[ ()]{}

\bibitem[Tokovinin \& Smekhov (2002)]{TS02}
Tokovinin, A. A. \& Smekhov, M. G. 2002 A\&A, 382, 118

\bibitem[Tokovinin et al. (2006)]{Tok06}
Tokovinin, A., Thomas, S., Sterzik, M., \& Udry, S.  2006, A\&A,  450, 681  



\bibitem[Tokovinin, Mason, \& Hartkopf (2010a)]{TMH10}
Tokovinin, A., Mason, B. D., \& Hartkopf, W. I. 2010a,  AJ, 139, 743

\bibitem[Tokovinin, Hartung, \& Hayward (2010b)]{THH10}
Tokovinin, A., Hartung, M., \& Hayward, Th. L. 2010b, AJ, 140, 510

\bibitem[Tokovinin (2011)]{ANDICAM}
Tokovinin, A.  2011, AJ, 141,  52 

\bibitem[Tokovinin et al. (2012)]{astrom1} 
Tokovinin, A., Hartung, M., Hayward, Th. L., \& Makarov, V. V.  2012, AJ,
144, 7 

\bibitem[Tokovinin et al. (2013)]{astrom2} Tokovinin, A., Hartung, M., \&
  Hayward, Th. L., 2013, AJ, 146, 8

\bibitem[Tokovinin \& L\'epine (2012)]{LEP}
Tokovinin, A. \& L\'epine, S. 2012, AJ, 144, 102 

%\bibitem[Tokovinin (2013)]{11072}
%Tokovinin, A. 2013, AJ, 145, 76

\bibitem[Tremko et al. (2010)]{Tremko2010}
Tremko, J., Bakos, G.A., Ziznivsky, J. et al. 2010, 
Contrib. Astron. Obs. Skalnate Pleso, 40, 83

\bibitem[van Leeuwen (2007)] {HIP2}
van Leeuwen, F. 2007, A\&A, 474, 653

\bibitem[Zacharias et al. (2005)]{NOMAD}
 Zacharias, N., Monet, D. G., Levine, S. E. et al. 2005, A\&AS, 204, 4815 
%Urban S.E., Gaume R., Wycoff G.L.    <San Diego AAS Meeting, January (2005)>
%    =2004AAS...205.4815Z


\end{thebibliography}
\end{document}